\documentclass[review,11pt,3p]{elsarticle}

\usepackage{amsmath,amsfonts,amssymb}
\usepackage{cases}
\usepackage[version=4]{mhchem}
\usepackage{amsthm}
\usepackage{bm}
\usepackage{newtxtext,newtxmath}
\usepackage{graphicx,epstopdf}
\usepackage[position=top,labelfont=normalfont,textfont=normalfont,singlelinecheck=off,justification=raggedright]{subfig}
\usepackage{floatrow}
\usepackage[dvipsnames]{xcolor}
\usepackage{setspace}
\usepackage{enumerate,etaremune}
\usepackage{booktabs}
\usepackage{tabularx,multirow}
\usepackage{framed}
\usepackage{hhline}
\usepackage{url}
\usepackage{import}
\usepackage[unicode,bookmarks=false]{hyperref}
\hypersetup{
  colorlinks,
  citecolor=Blue,linkcolor=Blue,urlcolor=Blue
}

\usepackage{lineno}

\usepackage[textsize=scriptsize,colorinlistoftodos,
  linecolor=Goldenrod,backgroundcolor=Goldenrod!20,bordercolor=Goldenrod]
  {todonotes}

\usepackage{tikz}
\usetikzlibrary{shapes.geometric}
\usetikzlibrary{shapes,arrows}
\usetikzlibrary{plotmarks}

\usetikzlibrary{calc}

\usepackage{physics}

\usepackage{algorithm}
\usepackage{algorithmicx,algpseudocode}
\usepackage{cases}

\newcommand{\eg}{{\it e.g.}}
\newcommand{\ie}{{\it i.e.}}

\newcommand{\tensor}[1]{\bm{#1}}
\newcommand{\stress}{\sigma}
\newcommand{\strain}{\varepsilon}

\newcommand{\jump}[1]{\lbrack\!\lbrack #1 \rbrack\!\rbrack}

\newcommand{\pd}{\partial}

\newcommand{\el}{\mathrm{e}}

\newcommand{\pore}{\mathrm{pore}}
\newcommand{\m}{\mathrm{mineral}}

\newcommand{\rn}[1]{\uppercase\expandafter{\romannumeral #1\relax}}

\DeclareMathOperator{\vol}{vol}

\newsavebox{\dotbox}

\theoremstyle{remark}
\newtheorem{remark}{Remark}

\newcommand{\revised}[1]{{\color{black} #1}}

\newcolumntype{L}[1]{>{\raggedright\let\newline\\arraybackslash\hspace{0pt}}m{#1}}
\newcolumntype{C}[1]{>{\centering\let\newline\\arraybackslash\hspace{0pt}}m{#1}}
\newcolumntype{R}[1]{>{\raggedleft\let\newline\\arraybackslash\hspace{0pt}}m{#1}}

\allowdisplaybreaks

\biboptions{compress,comma,authoryear}
\AtBeginDocument{\hypersetup{citecolor=MidnightBlue,linkcolor=MidnightBlue,urlcolor=MidnightBlue}}

\usepackage{etoolbox}
\makeatletter
\patchcmd{\ps@pprintTitle}
{Preprint submitted to}
{~}
{}{}
\makeatother

\begin{document}

\begin{frontmatter}

\title{A Poromechanics-Based Framework for Fully Coupled Reactive Transport and Geomechanics in Porous Rocks}
\address[LLNL-AEED]{Atmospheric, Earth, and Energy Division, Lawrence Livermore National Laboratory, USA}
\address[LLNL-CED]{Computational Engineering Division, Lawrence Livermore National Laboratory, USA}
\address[NWU]{Department of Civil and Environmental Engineering, Northwestern University, USA}
\address[EarthFlowAI]{EarthFlow AI}

\author[LLNL-AEED]{Fan Fei\corref{corr}}
\cortext[corr]{Corresponding authors}
\ead{fei2@llnl.gov}

\author[NWU]{Yifan Yang\corref{corr}}
\ead{yifanyang2026@u.northwestern.edu}

\author[LLNL-AEED]{Aleksander L. Zibitsker}

\author[NWU]{Giuseppe Buscarnera}

\author[LLNL-AEED,EarthFlowAI]{Randolph R. Settgast}

\author[LLNL-AEED]{Kari Finstad}

\author[LLNL-CED]{Giovanna Bucci}

\journal{~}

\begin{abstract}

Mineral precipitation and dissolution in rocks alter pore structure, stress, and hydraulic properties, producing tightly coupled chemo-hydro-mechanical processes that are central to subsurface systems.
Yet, existing continuum-scale simulators lack a generalizable, mathematically tractable, and physically grounded chemo-mechanics formulation for modeling these coupled processes.
To address this gap, this work presents a poromechanics-based framework for chemo-mechanics coupling and integrates it into coupled reactive transport and geomechanics simulations.
Building upon classical poromechanics theory, the framework provides distinct descriptions for the stress states of the host rock and pore minerals, \revised{in addition to pore fluid pressure,} enabling a rigorous and flexible treatment of their individual behavior and mechanical interactions during precipitation and dissolution.
As demonstrated by numerical examples, the poromechanics-based method captures the expected mechanical response by translating pore-scale mineral growth into mineralization pressure.
By accounting for mineral compressibility, the method also supports more physically realistic representations of deformation and induced stress.
\revised{Numerical comparisons} further highlight the unique capability of the poromechanics-based approach to model effective stress evolution and fracture development without explicitly representing pore geometry or other microstructural features.
Overall, this framework offers a versatile, physics-based predictive approach to modeling coupled chemo-mechanical processes in reactive porous rocks and supports future reservoir-scale analyses of engineered subsurface systems.

\end{abstract}

 
\end{frontmatter}


\section{Introduction}
\label{sec:intro}

The precipitation and dissolution of minerals in the pores of geologic materials are fundamental geochemical processes governing the evolution of subsurface systems across a wide range of natural and engineered environments. 
In particular, these processes play a critical role in applications such as geologic carbon sequestration, oil and gas recovery, geothermal energy production, and critical mineral extraction~\citep{oelkers2008mineral,taron2009thermal,gislason2014carbon,pyrak2015controlling,martens2021toward,omar2021co,clavijo2022coupled,neil2024integrated,nisbet2024carbon}, where fluid--rock interactions continuously modify the mechanical and physical states of the host formation, thereby influencing reservoir performance and long-term stability.

The complexity of reaction-induced evolution stems from the inherently multiscale and multiphysics nature of mineral precipitation and dissolution.
At the pore scale, these reactions cause morphological changes that modify pore structure and drive deformation in both the host rock skeleton and mineral phases within the pores, resulting in intricate mechanical behavior in each component~\citep{Dormieux2006microporomechanics,yang2025rockmech} as well as complex interactions between them~\citep{mura1987micromechanics,xu2025feedback}.
These microscale alterations manifest at the macroscopic scale as the evolution of rock porosity and permeability that control fluid transport efficiency.
More importantly, the volumetric changes associated with these alterations induce stress redistribution within the rock matrix, which can significantly affect the mechanical integrity of the subsurface system and, in turn, modulate subsequent reactive transport processes~\citep{kelemen2012reaction,zhu2016experimental,xing2018generating,renard2021reaction,uno2022volatile,nisbet2024carbon}.
One representative example is serpentinization, in which hydration of olivine produces volume expansion of the rock matrix by up to 50\%~\citep{macdonald1985rate}.
When the resulting stress induced by volume expansion exceeds the tensile strength of the host rock, reaction-driven fractures form and provide additional fluid pathways and mineral surface areas for reaction~\citep{kelemen2012reaction,zhu2016experimental,xing2018generating,zheng2019mixed,renard2021reaction}.
Conversely, when the induced stress remains low, mineral precipitation can clog pore space and impede fluid flow~\citep{wolterbeek2018reaction,zheng2019mixed,renard2021reaction}.
This intricate coupling among chemical reactions, mechanical deformation, and fluid transport across scales necessitates \revised{comprehensive} numerical frameworks for reliable prediction and operational planning in subsurface systems.

Over the past decades, several numerical codes and software platforms have been developed to couple reactive transport and geomechanics.
Representative examples for reservoir-scale simulations include TOUGHREACT--FLAC3D~\citep{rutqvist2011status}, OpenGeoSys~\citep{kolditz2012opengeosys}, RetrasoCodeBright~\citep{kvamme2009reactive}, and FALCON~\citep{podgorney2021reference}, which generally integrate a reactive transport code with a dedicated geochemistry module (\eg~TOUGHREACT~\citep{xu2011toughreact}, PHREEQC~\citep{parkhurst2013phreeqc}) into a geomechanics code (\eg~FLAC3D~\citep{itasca2011v5}, CODE\_BRIGHT~\citep{olivella1996codebright}, MOOSE~\citep{permann2020moose}, etc.) in a sequential manner.
These platforms have been extensively applied to the geothermal reservoirs~\citep{taron2009thermal,podgorney2021reference}, carbon sequestration~\citep{kvamme2009reactive,xiao2020chemical,clavijo2022coupled,mura2024integration}, nuclear waste disposal~\citep{rutqvist2014modeling,zheng2015impact,montoya2022ogsdgd}.
At the pore scale, discrete modeling approaches have also been employed to explicitly resolve grain-scale mechanics and fluid-solid interactions during mineral reactions~\citep{yoshida2020fluid,wu2025unraveling}.

Despite these advances, most existing formulations either neglect the chemo-mechanical coupling at the constitutive level or represent it through simplified empirical or phenomenological relationships, rather than rigorously resolving poromechanical interactions from first principles.
For instance, many studies directly scale rock mechanical properties (\eg~elastic moduli) as functions of mineral volume fractions or porosity changes using empirical degradation factors~\citep{pignatelli2013coupled,schuler2020chemo,clavijo2022coupled,guo2024reactive,mollaali2025variational,ventura2026phase}.
Another class of approaches introduces chemical alteration through chemoplasticity~\citep{hueckel2002reactive,hu2013environmentally}, where chemical variables such as species concentrations/activities or dissolved/precipitated mineral mass enter the elastoplastic constitutive law by modifying cohesion, friction angle, hardening variables, or the size and shape of the yield surface.
Although these formulations can capture important processes such as reaction-assisted softening and cracking, particularly under dissolution, they lack the generality to extrapolate beyond the specific materials and conditions for which they are developed and calibrated.
Also notably, the porosity update is often over-simplified, being commonly assumed to depend solely on mineral volume fraction and/or fluid pressure~\citep{nghiem2004modeling,rutqvist2014modeling,hao2019multiscale,smith2019validation,guo2023computational,mura2024integration}, without sufficiently accounting for the compressibility of the pore structure and the fluid and mineral solids therein.

\revised{One common strategy for continuum-based chemo-mechanical coupling is to treat the volumetric changes associated with mineral precipitation and dissolution as an inelastic strain, emulating the classical eigenstrain method of micromechanics~\citep{mura1987micromechanics,pichler2010eigenstrain,xu2025feedback}.}
\revised{Due to its conceptual simplicity and straightforward implementation within existing mechanics solvers, this strategy has been widely applied across diverse chemo-mechanical problems, including serpentinization~\citep{evans2018poroelastic,evans2020phase,malvoisin2021achieving}, clay swelling~\citep{laredj2010modeling,rutqvist2014modeling,zheng2017coupled}, and also lithium-ion batteries~\citep{bucci2014measurement,bucci2016formulation,narayan2020modeling,uzun2025insights,li2026coupled,lin2026mechanics}.}
\revised{This simplicity, however, rests on a key assumption.
Unlike the classical eigenstrain approach~\citep{pichler2010eigenstrain,xu2025feedback}, in which the influence tensor\footnote{\revised{In the classical eigenstrain theory, the influence tensor is defined as a fourth-order linear operator that maps the eigenstrain distributed through a body or a constituent phase to the induced local stress or strain. It usually depends on the stiffness and geometry of all constituent phases.}} is formulated separately from the homogenized stiffness of the whole system, the inelastic-strain treatment above relies on a single stiffness tensor for the whole system, implicitly assuming that the pore-filling minerals and the host rock skeleton share the same elastic properties.}
\revised{As a result, the method fails to capture the poromechanical nature of geomaterials, in which mineral deformation and pore structure evolution of the host rock are governed by coupled yet distinct physical laws.}
\revised{Although distinct elastic stiffness could in principle be considered for the pore-filling minerals, the derivation of the influence tensor becomes increasingly difficult when fluid and multiple minerals coexist in the pore space.}
\revised{Alternatively, the microstructure can be explicitly resolved to separately capture the host rock, the pore-filling minerals, and the fluid, but this is computationally infeasible at the scales relevant to geologic systems.}

To pursue a generalizable, physically grounded description of chemo-mechanical coupling in porous geomaterials, this study builds on a poromechanics-based analytical framework recently proposed by \citet{yang2025poromechanical} and integrates it into coupled reactive transport and geomechanics simulation.
\revised{Unlike the inelastic-strain-based method, this framework explicitly distinguishes the stress states of the host rock and the pore-filling minerals, in addition to the pore fluid pressure, thereby allowing for a more rigorous and tractable formulation to capture the mechanical interactions among these phases during precipitation and dissolution.}
The porosity update likewise follows from the poromechanical response, accounting for the deformation of the pore space and the compressibility of the solid and fluid within it.
Another key feature of the poromechanics-based framework is its ability to bridge scales, translating pore-scale mineral growth into a pore mineral pressure acting on the host rock, so that mineralization-induced mechanical responses can be captured at the meso- and reservoir-scale without explicitly resolving the pore geometry.
In the same poromechanics spirit, \citet{choo2018cracking} also developed a chemo-hydro-mechanical formulation that represented crystallization pressure as a separate internal mechanical loading, but it did not include reactive transport coupling.
Overall, by reconciling first-principles poromechanics with reactive transport, this work establishes a more rigorous foundation for assessing the coupled chemo-mechanical processes in reactive subsurface systems.

\revised{Both the poromechanics-based and the inelastic-strain-based methods are implemented in the open-source multiphysics simulator GEOS~\citep{settgast2024geos}, where the mechanics module is sequentially coupled with a single-phase reactive transport module.
For the latter, we adopt the formulation of \citet{evans2018poroelastic,evans2020phase}, which provides a general means of computing the reaction-induced inelastic strain directly from mineral reaction rates.}
For simplicity, the present implementation is restricted to isotropic material behavior.
Nevertheless, the poromechanics-based framework is readily extensible to a full tensorial form for future anisotropic generalizations.
Together, these components form a fully coupled chemo-hydro-mechanical solver, in which chemical reactions dynamically influence mechanical behavior and, in turn, mechanical changes \revised{(\eg~rock fracture)} feed back into transport properties \revised{such as permeability and diffusivity}.

The remainder of this paper is organized as follows.
Section~\ref{sec:equations} provides the governing equations for coupled geomechanics and reactive transport processes.
Section~\ref{sec:coupling} introduces the poromechanics-based framework for chemo-mechanical coupling, which constitutes the core contribution of this work.
\revised{Analytical comparisons between the poromechanics-based and the inelastic-strain-based approaches under idealized conditions are also provided in this section.}
Section~\ref{sec:numerical} describes the numerical discretization and solution algorithm.
\revised{Section~\ref{sec:results} presents representative simulations and demonstrates the key features of the poromechanics-based approach through corresponding numerical comparisons.}
Finally, Section~\ref{sec:closure} concludes the work and discusses future research directions.

\section{Problem description and governing equations}
\label{sec:equations}

\begin{figure}[htbp]
    \centering
    \includegraphics[width=\textwidth]{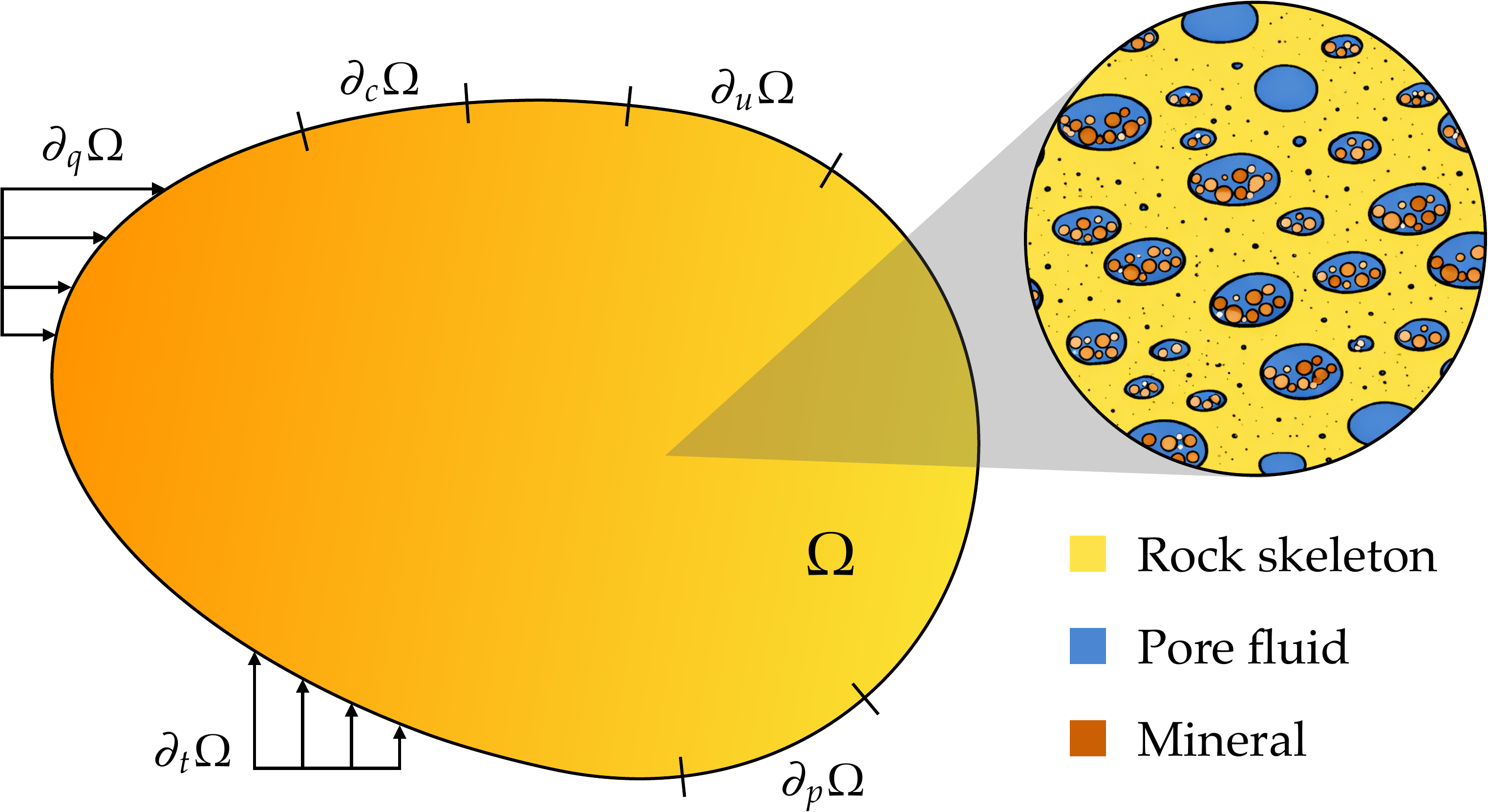}
    \caption{Problem description of a coupled chemo-hydro-mechanical system in porous rock matrix.}
    \label{fig:problemStatement}
\end{figure}

This section provides the governing equations for modeling of coupled geomechanics and reactive transport processes. 
Figure~\ref{fig:problemStatement} presents a schematic of a coupled chemo-hydro-mechanical system, where mineral precipitation and dissolution occur within the pore space of rock matrix and result in the mechanical response. 
Here, we consider the problem domain denoted by $\Omega \in \mathbb{R}^{n_\mathrm{dim}}$, where $n_\mathrm{dim}$ means the spatial dimension of interest.
The external boundary $\pd \Omega$ is partitioned into complementary subsets, \ie~$\pd_{u} \Omega \cup \pd_{t} \Omega = \pd_{p} \Omega \cup \pd_{q} \Omega$ and $\pd_{u} \Omega \cap \pd_{t} \Omega = \pd_{p} \Omega \cap \pd_{q} \Omega = \emptyset$, where $\pd_{u} \Omega$ and $\pd_{t} \Omega$ are the boundaries with prescribed displacement and traction, respectively, while $\pd_{p} \Omega$ and $\pd_{q} \Omega$ are the boundaries with prescribed pore pressure and flux, respectively.
The remaining boundary condition $\pd_{c} \Omega$ denotes the Dirichlet boundary for the concentration of chemical species.
For this coupled problem, we solve for the primary variables at time $t\in(0, t_{\max}]$, including the displacement ${u}_{i}$ with $i$ the index of spatial dimension, fluid pore pressure $p_f$, and the concentration of primary species $C_{s}$, where $s$ denotes the index for the primary species.
These primary variables are solved by the governing equations, which consist of the momentum balance equation, mass balance equations for the pore fluid and each chemical species, respectively, whose detailed formulations are given below.

\begin{remark}
Note that the primary species concentration is solved in the natural logarithmic form, \ie~$\ln C_{s}$, in the numerical implementation.
The reason behind this choice is to ensure the positivity of the concentration and to enhance the numerical stability of the nonlinear solver.
For simplicity, the concentration is expressed in molarity ($\mathrm{mol/m}^3$) throughout the formulation, unless otherwise specified.
The model implementation by default takes species concentration input in molality ($\mathrm{mol/kg}$), which is internally converted to molarity via the solvent density under the dilute-solution assumption, \ie~molarity $=$ molality $\times \, \rho_{w}$, where $\rho_{w}$ is the density of water.
\end{remark}

\subsection{Governing equations for momentum balance and fluid mass balance}
At the continuum scale, the governing equations describing the hydro-mechanical behavior of porous rock matrix are adopted from \citet{coussy2004poromechanics} and \citet{Lewis1998poromechanics}. 
The formulation is sufficiently general to accommodate the subsequent incorporation of chemical governing equations and chemo-mechanical coupling effects. 
Given the requirements of the numerical framework employed in this study, a Lagrangian description is adopted with small strain assumption. 
Accordingly, the stress, strain, and matrix porosity used in the following derivations correspond to Cauchy stress, infinitesimal strain, and Lagrangian porosity, respectively. 

To begin, the generic momentum balance equation is written as
\begin{align}
    \dfrac{\pd \stress_{ij}}{\pd x_j} + \rho g_{i} = 0, \label{eq:momentum-balance}
\end{align}
where $\stress_{ij}$ is the total stress of the rock matrix system,
$\rho$ represents the total density, and $g_{i}$ is the gravitational vector. 
For simplicity, the body force is ignored here, \ie~$\rho g_i = 0$.
Equation~\eqref{eq:momentum-balance} is subject to the boundary and initial conditions,
\begin{alignat}{2}
    u_{i} &= \bar{u}_{i} &\quad \text{on } \, &\pd_{u} \Omega, \\
    \stress_{ij} n_{j} &= \bar{t}_{i} &\quad \text{on } \, &\pd_{t} \Omega, \\
    u_{i}(x_j, 0) &= u_{i,0}(x_j) &\quad \text{in } \, &\Omega,
\end{alignat}
where $\bar{u}_{i}$ and $\bar{t}_{i}$ are the prescribed displacement and traction, respectively, $n_{j}$ is the outward unit normal vector to $\pd \Omega$, and $u_{i,0}$ is the prescribed initial displacement.

In parallel, the mass balance equation for the pore fluid is given as,
\begin{align}
    \dfrac{\partial \left( \phi \rho_{f} \right)}{\partial t}  + \dfrac{\partial\left( \rho_{f} v_{i} \right) }{\partial x_{i} } = \mathrm{sink/source},  \label{eq:flow-mass-balance}
\end{align}
where $\phi$ is the porosity, $\rho_{f}$ is the fluid density, and $v_{i}$ is the fluid velocity relative to rock skeleton.
According to the Darcy's law, $v_{i}$ can be evaluated as 
\begin{align}
    v_{i} = -\dfrac{k_{ij}}{\mu_{f}} \left(\dfrac{\pd p_{f}}{ \pd x_{j}} - \rho_{f} g_{j} \right), 
\end{align}
where $k_{ij}$ is the matrix permeability tensor, and $\mu_{f}$ is the fluid viscosity.
While advanced permeability models (\eg~the Carman--Kozeny model~\citep{kozeny1927,carman1937}) can be adopted, the permeability is assumed to be isotropic and constant throughout this work unless otherwise specified, \ie~$k_{ij} = k_0 \delta_{ij}$, where $k_0$ is the reference permeability.
Equation~\eqref{eq:flow-mass-balance} is subject to the boundary and initial conditions,
\begin{alignat}{2}
    p_{f} &= \bar{p}_{f} &\quad \text{on } \, &\pd_{p} \Omega, \\
    -\rho_{f} v_{i} n_{i} &= \bar{q} &\quad \text{on } \, &\pd_{q} \Omega, \\
    p_{f}(x_j, 0) &= p_{f,0}(x_j) &\quad \text{in } \, &\Omega,
\end{alignat}
where $\bar{p}_{f}$ and $\bar{q}$ are the prescribed pore pressure and fluid mass flux, respectively, and $p_{f,0}$ is the prescribed initial pore pressure.

The formulations of $\stress_{ij}$ and $\phi$ are detailed further in Section~\ref{sec:coupling}, where the chemo-mechanical coupling is introduced.

\subsection{Governing equations for mass balance of chemical species}
The remaining equation sets govern the mass balance of chemical species, whose general form is given below,
\begin{align}
    \dfrac{\pd \phi C^\mathrm{total}_{s} }{\pd t} + \dfrac{\pd C^\mathrm{total}_{s} v_{i}}{\pd x_{i}} - \dfrac{\pd \phi_0 \mathcal{D} (\pd C^\mathrm{total}_{s}/\pd x_{i}) }{\pd x_i} = \text{species sink/source} - \sum^{n_\mathrm{react}}_{m} \nu_{s, m} R_{m}, \label{eq:species-mass-balance}
\end{align}
where $C^\mathrm{total}_{s}$ is the total/aggregate molarity for the primary species $s$, $\phi_0$ is the reference (initial) porosity, $\mathcal{D}$ is the diffusion/dispersion coefficient, assumed identical for all species, $\nu_{s, m}$ is the stoichiometric coefficient for the species $s$ at the mineral reaction $m$, and $R_{m}$ is the molarity rate for the mineral reaction $m$, taken positive for precipitation.
Equation~\eqref{eq:species-mass-balance} is subject to the boundary and initial conditions
\begin{alignat}{2}
    C_{s} &= \bar{C}_{s} &\quad \text{on } \, &\pd_{c} \Omega, \\
    C_{s}(x_j, 0) &= C_{s,0}(x_j) &\quad \text{in } \, &\Omega,
\end{alignat}
where $\bar{C}_{s}$ is the prescribed species concentration, and $C_{s,0}$ is the prescribed initial species concentration.

Specifically, the total molarity of the primary species is calculated as
\begin{align}
    C^\mathrm{total}_{s} := C_{s} + \sum_{r}^{n_\mathrm{eq}} \nu_{s, r} C^\mathrm{sec}_{r},
\end{align}
where $C^\mathrm{sec}_{r}$ is the concentration of the secondary species in its corresponding equilibrium reaction $r$, $\nu_{s,r}$ is the stoichiometric coefficient for the primary species $s$ in the equilibrium reaction $r$. 
The secondary species concentration is calculated as
\begin{align}
    C^\mathrm{sec}_r = \dfrac{K^\mathrm{eq}_r}{\gamma^\mathrm{sec}_r}\prod_{s}(\gamma^\mathrm{prim}_s C_s)^{\nu_{s, r}}, 
\end{align}
where $K^\mathrm{eq}_r$ is the equilibrium constant of the reaction $r$, $\gamma^\mathrm{prim}_{s}$ and $\gamma^\mathrm{sec}_{r}$ are activity coefficients for the primary and secondary species, respectively. 
The activity coefficient for each species is obtained from the adopted activity model, \eg~B-dot model~\citep{helgeson1969thermodynamics}. 
For simplicity, we consider all activity coefficients to be unity in the current work, \ie~$\gamma^\mathrm{prim}_s = \gamma^\mathrm{sec}_r = 1$.
The molarity rate for the mineral reaction $m$ can be computed as 
\begin{align}
    R_m = A_{m}k_{m}\left(\dfrac{Q_m}{K^\mathrm{eq}_{m}} - 1\right), \label{eq:reaction-rate}
\end{align}
where $A_{m}$ is the specific surface area of the mineral, $k_{m}$ is the rate constant that depends on the temperature and acidity, $K^\mathrm{eq}_{m}$ is the equilibrium constant for the mineral reaction, and $Q_{m}$ is the reaction quotient that can be obtained as
\begin{align}
    Q_m := \prod_{s} (\gamma^\mathrm{prim}_{s} C_{s})^{\nu_{s, m}}.
\end{align}
All total molarity and kinetic reaction rates are evaluated in our recently developed open-source library HPCReact~\citep{settgast2025hpcreact}, which is designed to be GPU-callable for modeling pointwise chemical reactions.

\section{Chemo-mechanics coupling}
\label{sec:coupling}

To close the governing equations for the momentum balance and mass balance in Section~\ref{sec:equations}, this section presents chemo-mechanics coupling that specifies the formulations of the total stress $\stress_{ij}$ and the porosity $\phi$.
\revised{We begin with a review of the inelastic-strain-based method, which is widely adopted in the literature to incorporate the mechanical effects of mineral reactions.
After identifying its key limitations, we introduce} a more general and flexible poromechanics-based coupling method rooted in the analytical framework of \citet{yang2025poromechanical}.

\subsection{\texorpdfstring{\revised{Inelastic-strain-based method and its limitations}}{Inelastic-strain-based method and its limitations}}
\revised{Formulations adopting the inelastic-strain-based method commonly share the strain decomposition below, in which the total strain rate is additively split into an elastic part and an inelastic part induced by mineral reactions~\citep{rutqvist2014modeling,zheng2017coupled,evans2018poroelastic,evans2020phase},
\begin{align}
    \dot{\strain}_{ij} = \dot{\strain}^{\el}_{ij} + \dot{\strain}^{\ast}_{ij}, \label{eq:inelastic-strain-decomposition}
\end{align}
where other inelastic strains, such as plastic and thermal strains, can be incorporated in the same manner.
While the specific expression of ${\strain}^{\ast}_{ij}$ differs across the literature, we take the formulation of \citet{evans2018poroelastic,evans2020phase} as a representative example, since it offers a general relation between $\dot{\strain}^{\ast}_{ij}$ and the mineral reaction rate.
For simplicity of calculation, their formulation assumes an isotropic inelastic strain,} $\dot{\strain}^{\ast}_{ij} = \dot{\bar{\strain}}^{\ast}\delta_{ij}$, where $\dot{\bar{\strain}}^{\ast}$ can be obtained as
\begin{align}
    \dot{\bar{\strain}}^{\ast} := \sum^{n_\mathrm{react}}_{m} \dfrac{R_m M_{m}}{\rho_{m}}.\label{eq:eigen-strain-evans}
\end{align}
Here, $M_{m}$ and $\rho_m$ are the molar weight and density of the mineral $m$, respectively.

\revised{With the formulation of the inelastic strain, the total stress is expressed as~\citep{rutqvist2014modeling,evans2018poroelastic,evans2020phase}},
\begin{align}
    \dot{\stress}_{ij} = \dot{\hat{\stress}}_{ij} -  \dot{p}_{f} \delta_{ij}, \,\, \text{with } \dot{\hat{\stress}}_{ij} = \mathbb{C}_{ijkl} \dot{\strain}^{\el}_{kl} = \mathbb{C}_{ijkl} \left(\dot{\strain}_{kl} - \dot{\bar{\strain}}^{\ast} \delta_{kl}\right), \label{eq:eigen-strain-total-stress-evans}
\end{align}
where $\hat{\stress}_{ij}$ is the (Terzaghi) effective stress of the solid skeleton, lumping together elastic deformation of the host rock and \revised{the mineral inelastic strain, while the pore fluid pressure $p_f$ is accounted for separately}.
Considering isotropy and linear elasticity for the host rock, we can formulate $\mathbb{C}_{ijkl}$ as
\begin{align}
    \mathbb{C}_{ijkl} := K\delta_{ij}\delta_{kl} + 2G \left(\mathbb{I}_{ijkl} - \dfrac{1}{3} \delta_{ij} \delta_{kl} \right), \label{eq:elastic-stiffness}
\end{align}
where $K$ and $G$ are the drained bulk modulus and the shear modulus of the rock, respectively, and $\mathbb{I}_{ijkl}$ is the fourth-order identity tensor.

\revised{Despite its simplicity, this formulation---specifically the stress calculation~\eqref{eq:eigen-strain-total-stress-evans} along with the strain decomposition~\eqref{eq:inelastic-strain-decomposition}---exhibits several key limitations:}
\begin{enumerate}[(i)]
    \item \revised{The stress induced by the inelastic strain is computed through a single stiffness tensor $\mathbb{C}_{ijkl}$, implicitly assuming that the mineral(s) produced/consumed in the pore space share the same elastic properties as the host rock, but they often differ in reality.}
    \item \label{item:limitation-constitutive-tensor} Even if a distinct mineral stiffness/compressibility is considered, \revised{it requires a separate operator to map the inelastic strain to the induced stress (\eg~the influence tensor in the classical eigenstrain theory), whose derivation depends on the microstructure and can be nontrivial} if multiple substances/phases coexist in the pore.
    \item \label{item:limitation-lumped-stress} \revised{The effective stress term in Eq.~\eqref{eq:eigen-strain-total-stress-evans} combines the host rock and the pore-filling minerals into a single solid phase, and thus cannot resolve the individual stress state of the host rock skeleton.
    {Accordingly, the positive inelastic strain caused by pore mineral growth produces net compression throughout the solid phase under confinement. 
    Consequently, mineral precipitation alone cannot induce tension-driven fracture at the material point where the reaction occurs, 
    except when material properties or reaction rates are heterogeneous.}  
    {In this case, the spatial heterogeneity} must be explicitly resolved at its characteristic length scale, ultimately posing challenges for simulations across scales.}
\end{enumerate}

\subsection{Proposed poromechanics-based approach}
Motivated by the limitations above, we provide a poromechanics-based approach to chemo-mechanics coupling.
Following the analytical framework proposed by \citet{yang2025poromechanical}, the generalized formulation for the macroscopic total stress of the system can be written as,
\begin{align}
    \dot{\stress}_{ij} = \dot{\stress}'_{ij} + \alpha_{ijkl} \dot{\stress}^\pore_{kl},\,\, \text{with } \dot{\stress}'_{ij} = \mathbb{C}_{ijkl}\dot{\strain}_{kl}  \label{eq:biot-total-stress}
\end{align}
where $\stress'_{ij}$ denotes the effective stress of the host rock skeleton itself and ${\stress}^\pore_{ij}$ is the overall stress exerted on the pore walls by all constituents residing in the pore space.
Note that unlike $\hat{\stress}_{ij}$ in Eq.~\eqref{eq:eigen-strain-total-stress-evans}, which lumps together the host rock and \revised{mineral inelastic strain} contributions, $\stress'_{ij}$ here refers to the host rock only\revised{, so that the proposed model can evaluate the rock's mechanical integrity locally based on its own stress state, without explicitly resolving the {spatial} heterogeneity\revised{, thereby addressing limitation~(\ref{item:limitation-lumped-stress})}}.
In addition, $\alpha_{ijkl}$ in Eq.~\eqref{eq:biot-total-stress} stands for the full fourth-order Biot coefficient tensor\footnote{The Biot coefficient quantifies how much of the internal pore stress (\eg~pore fluid pressure) contributes to the pore deformation and total stress of the porous medium. \revised{The tensorial form further accounts for anisotropy arising from complex pore structures and behavior of pore-filling materials.}}.
As derived in \citet{yang2025poromechanical}, it is defined based on the rock skeleton stiffness $\mathbb{C}_{ijkl}$ and rock grain\footnote{The rock grains are the solid particle constituents that bond with each other to form the porous skeleton of the rock.} stiffness ${\mathbb{C}^{\mathrm{s}}_{ijkl}}$ as,
\begin{align}
     \alpha_{ijkl} := {\left(\mathbb{C}^{\mathrm{s}}_{ijmn}\right)^{-1}} \mathbb{C}_{mnkl} - \mathbb{I}_{ijkl}
\end{align}
It is clear that the above relation can be reduced to the definition of a scalar Biot coefficient below in the classical Biot poromechanics formulation,
\begin{align}
    \alpha := 1 - \dfrac{K}{K^{\mathrm{s}}},
\end{align}
where $K^{\mathrm{s}}$ represents the bulk modulus of the rock solid grain. 
In addition, the generalized formulation of the second-order porosity tensor is given as
\begin{align}
    \dot{\phi}_{ij} = \phi_{0}\dot{\strain}^{\phi}_{ij} = \alpha_{ijkl} \dot{\strain}_{kl} + \gamma_{ijkl} \dot{\stress}^\pore_{kl},
\end{align}
where $\gamma_{ijkl}$ denotes the fourth-order tensor for the confined storage coefficient, which is defined as
\begin{align}
    \gamma_{ijkl} := -{\left(\mathbb{C}^{\mathrm{s}}_{ijmn}\right)^{-1}} \left(\alpha_{mnkl} - \phi_{0} \mathbb{I}_{mnkl} \right) .
\end{align}

A standout advantage of this poromechanics-based method over the \revised{inelastic-strain-based method} is \revised{that} it adds more flexibility to account for the mechanical feedback associated with mineralization. 
More specifically, even though the \revised{inelastic strains} of different materials in the pore can be superimposed, the derivation of \revised{the corresponding operator that maps these strains to the induced stress} poses \revised{a} significant challenge\revised{, as noted in limitation~(\ref{item:limitation-constitutive-tensor})}. 
On the other hand, this poromechanics-based method allows the overall pore stress $\dot{\stress}^\pore_{ij}$ to be decomposed, for example, additively into, 
\begin{align}
    \dot{\stress}^{\pore}_{ij} = \dot{\bar{\stress}}^{\m}_{ij} - \dot{p}_{f} \delta_{ij} = \sum^{n_{\m}}_{m} \dot{{\stress}}^{m}_{ij} - \dot{p}_{f} \delta_{ij}. \label{eq:pore-stress}
\end{align}
Here, we define a lump-sum mineral stress $\bar{\stress}^{\m}_{ij}$ as the summation of each individual stress contribution $\stress^{m}_{ij}$ from all minerals in the pore, which is tentatively given as,
\begin{align}
    \dot{{\stress}}^{m}_{ij} = {\mathbb{C}}^{m}_{ijkl} \left[ S^{m} (\dot{\phi}_{kl}/\phi_0 - \dot{p}_{f} \delta_{kl}/(3K^{f})) - \left(\dot{\strain}^\ast\right)^{m}_{kl}\right],
\end{align}
and therefore,
\begin{align}
    \dot{\bar{\stress}}^{\m}_{ij} = \sum^{n_\m}_{m} {\mathbb{C}}^{m}_{ijkl} S^{m} \left[\dot{\phi}_{kl}/\phi_0 - \dot{p}_{f} \delta_{kl}/(3K^{f})\right] - \sum^{n_\m}_{m} {\mathbb{C}}^{m}_{ijkl} \left(\dot{\strain}^\ast\right)^{m}_{kl} ,
    \label{eq:overall-mineral-stress}
\end{align}
where ${\mathbb{C}}^{m}_{ijkl}$ and $\left(\dot{\strain}^\ast\right)^{m}_{ij}$ are the stiffness and inelastic strain due to solid growth for the $m^\mathrm{th}$ mineral, respectively, $K^{f}$ is the bulk modulus of the pore fluid (\ie~the reciprocal of the fluid compressibility, $K^{f} = 1/c^{f}$), and $S^{m}$ is the saturation of the $m^\mathrm{th}$ mineral in the pore and satisfies $\sum^{n_{\m}}_{m} S^{m} = 1$.
The factor of 3 accompanying $K^{f}$ converts the fluid volumetric strain into the isotropic strain tensor.
For simplicity of experimental calibration, we further assume an identical stiffness for all minerals in the pore, \ie~$\mathbb{C}^{m}_{ijkl} = \bar{\mathbb{C}}^{\m}_{ijkl}$, so Eq.~\eqref{eq:overall-mineral-stress} reduces to
\begin{align}
    \dot{\bar{\stress}}^{\m}_{ij} = \bar{\mathbb{C}}^{\m}_{ijkl} \left[\dot{\phi}_{kl}/\phi_0 - \dot{p}_{f} \delta_{kl}/(3K^{f}) - \sum^{n_\m}_{m} \left(\dot{\strain}^\ast\right)^{m}_{kl}\right] . \label{eq:overall-mineral-stress-simplified}
\end{align}
Although more comprehensive formulations of the inelastic strain have been developed to relate solute molarity to crystal growth~\citep{yang2026correns}, \ie~the strain induced by supersaturation-driven mineral precipitation, the simpler \revised{yet general} correlation proposed by \citet{evans2018poroelastic,evans2020phase} is {chosen} in the present framework to facilitate a direct comparison on the mechanical {response predicted by} the \revised{inelastic-strain-based method} and the proposed poromechanics-based formulation. 
The adopted relationship is expressed as follows,
\begin{align}
    \left(\dot{\strain}^\ast\right)^{m}_{ij} = \left(\dot{\strain}^\ast\right)^{m} \delta_{ij} = \dfrac{R_m M_m}{\rho_m} \delta_{ij}.
\end{align}
Recalling Eq.~\eqref{eq:eigen-strain-evans} for the definition of $\dot{\bar{\strain}}^\ast$, the formulation of the overall mineral stress can be reduced to
\begin{align}
    \dot{\bar{\stress}}^{\m}_{ij} = \bar{\mathbb{C}}^{\m}_{ijkl} \left[\dot{\phi}_{kl}/\phi_0 - \dot{p}_{f} \delta_{kl}/(3K^{f}) - \dot{\bar{\strain}}^\ast \delta_{kl}\right] .
\end{align}

To further simplify numerical implementation, we assume an isotropic mineral stress and porosity, such that
\begin{align}
    \dot{\bar{\stress}}^{\m}_{ij} = - \dot{\bar{p}}_{\m} \delta_{ij} = \bar{K}^{\m} \left( \dot{\phi} / \phi_{0} - \dot{p}_{f}/K^{f} - 3\dot{\bar{\strain}}^\ast \right) \delta_{ij} .\label{eq:isotropic-mineral-stress-poromech}
\end{align}
Here, $\bar{p}_{\m}$ denotes the lump-sum pore mineral pressure, $\bar{K}^{\m}$ is the overall bulk modulus for all mineral phases in the pore. 
Combining Eqs.~\eqref{eq:biot-total-stress}, \eqref{eq:pore-stress}, \eqref{eq:overall-mineral-stress-simplified}, and \eqref{eq:isotropic-mineral-stress-poromech}, we can obtain the final formulation of the total stress as 
\begin{align}
    \dot{\stress}_{ij} = \mathbb{C}_{ijkl} \dot{\strain}_{kl} - \alpha \left( \dot{p}_{f} + \dot{\bar{p}}_{\m} \right) \delta_{ij} . \label{eq:poromech-final-stress}
\end{align}
Similarly, the simplified isotropic porosity can be formulated as
\begin{align}
    \dot{\phi} = \alpha\mathrm{tr}(\dot{\strain}_{ij}) + \dfrac{\alpha - \phi_{0}}{K^{\mathrm{s}}} \left( \dot{p}_{f} + \dot{\bar{p}}_{\m} \right)
    = \alpha\dfrac{\dot{\stress}'_{\vol}}{K}+ \dfrac{\alpha - \phi_{0}}{K^{\mathrm{s}}} \left( \dot{p}_{f} + \dot{\bar{p}}_{\m} \right), \label{eq:isotropic-porosity-poromech}
\end{align}
where $\dot{\stress}'_{\vol}$ is the volumetric part of the effective stress.
Inserting $\dot{\bar{p}}_{\m}$ from Eq.~\eqref{eq:isotropic-mineral-stress-poromech} into the above equation and rearranging the result leads to
\begin{align}
    \dot{\phi} = \left[\alpha\dfrac{\dot{\stress}'_{\vol}}{K} + \dfrac{\alpha - \phi_{0}}{K^{\mathrm{s}}} \left(\dfrac{\bar{K}^{\m}+K^{f}}{K^{f}}\dot{p}_{f} + 3\bar{K}^{\m} \dot{\bar{\strain}}^\ast \right) \right] / \left[ 1 + \dfrac{\alpha - \phi_{0}}{K^{\mathrm{s}}}\dfrac{\bar{K}^{\m} }{\phi_0}\right]. \label{eq:isotropic-porosity-poromech-expanded}
\end{align}
Note that the porosity $\dot{\phi}$ formulated in Eq.~\eqref{eq:isotropic-porosity-poromech} characterizes the evolution of the total pore space driven by skeleton deformation and pore pressurization, yet it does not directly account for the infilling of pore space by precipitated minerals.
Since the pore space is shared by both pore fluid and newly grown minerals, the pore storage capacity of the fluid is effectively reduced by the volumetric rate of mineral precipitation.
Accordingly, the effective fluid porosity rate entering the fluid mass balance equation is given as
\begin{align}
    \dot{\phi}_\mathrm{eff} = \dot{\phi} - \sum^{n_\mathrm{react}}_{m} \dot{\theta}_{m}, \label{eq:effective-fluid-porosity}
\end{align}
where $\dot{\phi}$ is updated from Eq.~\eqref{eq:isotropic-porosity-poromech}, and $\dot{\theta}_{m}:= R_m M_m/\rho_m$ is the rate of volume fraction change for mineral $m$.
Substituting $\phi_\mathrm{eff}$ for $\phi$ in the accumulation terms of Eqs.~\eqref{eq:flow-mass-balance} and \eqref{eq:species-mass-balance}, the modified mass balance equations can be written as
\begin{align}
    \dfrac{\partial \left( \phi_\mathrm{eff}\rho_{f} \right)}{\partial t}  + \dfrac{\partial\left( \rho_{f} v_{i} \right) }{\partial x_{i} } = \mathrm{sink/source},  \label{eq:flow-mass-balance-modified}
\end{align}
and
\begin{align}
    \dfrac{\pd \phi_\mathrm{eff} C^\mathrm{total}_{s} }{\pd t} + \dfrac{\pd C^\mathrm{total}_{s} v_{i}}{\pd x_{i}} - \dfrac{\pd \phi_0 \mathcal{D} (\pd C^\mathrm{total}_{s}/\pd x_{i}) }{\pd x_i}  = \text{species sink/source} - \sum^{n_\mathrm{react}}_{m} \nu_{s, m} R_{m}. \label{eq:species-mass-balance-modified}
\end{align}

\begin{remark}
{It is noted that we assume the elastic stiffness $\mathbb{C}_{ijkl}$~\eqref{eq:elastic-stiffness} to be held constant as porosity evolves.
This is because porosity alteration contributes only marginally to the overall stiffness change, whereas pore adaptivity (\eg~separation of slim pores), which is beyond the scope of this work, generally plays a much more significant role~\citep{yang2025rockmech}.}
\end{remark}

\subsection{Analytical comparison of mechanical response}
\label{sec:analytical-comparison}
The formulations of the total stress for both the \revised{inelastic-strain-based} and poromechanics-based methods are summarized below, with the assumption of all isotropic properties and constitutive behavior,
\begin{subnumcases}{}
    \dot{\stress}^\mathrm{inel}_{ij} = \mathbb{C}_{ijkl}:(\dot{\strain}_{kl} - \dot{\bar{\strain}}^\ast \delta_{kl}) - \dot{p}_{f} \delta_{ij} & \text{\revised{for the inelastic-strain-based method,}} \label{eq:stress_eigen} \\[1em]
    \dot{\stress}_{ij}^\mathrm{poro} = \mathbb{C}_{ijkl}:\dot{\strain}_{kl} - \alpha \left( \dot{p}_{f} + \dot{\bar{p}}_{\m} \right) \delta_{ij} & \text{for the poromechanics-based method.} \label{eq:stress_poro}
\end{subnumcases}
For simplicity of analytical comparison, we consider no excessive pore pressure increment (\ie~$\dot{p}_{f} = 0$) and the rock grain is incompressible (\ie~$K^{\mathrm{s}} \to \infty$, so $\alpha=1$).
Then, the above equations can be further reduced to
\begin{subnumcases}{}
    \dot{\stress}^\mathrm{inel}_{ij} = \mathbb{C}_{ijkl}:(\dot{\strain}_{kl} - \dot{\bar{\strain}}^\ast \delta_{kl}) & \text{\revised{for the inelastic-strain-based method,}} \label{eq:stress_eigen_simplified} \\[1em]
    \dot{\stress}_{ij}^\mathrm{poro} = \mathbb{C}_{ijkl}:\dot{\strain}_{kl} - \dot{\bar{p}}_{\m}  \delta_{ij} & \text{for the poromechanics-based method.} \label{eq:stress_poro_simplified}
\end{subnumcases}
With the simplified formulations of the total stress, we consider two extreme conditions below, namely, a zero stress or no confinement condition (\ie~$\dot{\stress}_{ij} = 0$), and a zero strain or full confinement condition (\ie~$\dot{\strain}_{ij} = 0$).
The latter can be viewed as an idealized limit of the real subsurface setting, where a reactive rock volume is well confined by the surrounding formation.

\paragraph{Scenario I – Zero stress condition}
This case simply considers 
\begin{align}
    \dot{\stress}_{ij} = \dot{\stress}^\mathrm{inel}_{ij}  = \dot{\stress}_{ij}^\mathrm{poro} = 0. 
\end{align}
To satisfy Eqs.~\eqref{eq:stress_eigen_simplified} and \eqref{eq:stress_poro_simplified} and inserting $\bar{p}_{\m}$ as formulated in Eq.~\eqref{eq:isotropic-mineral-stress-poromech}, we can obtain the resulting volumetric strain $\strain_{\vol}$ for both methods as
\begin{subnumcases}{}
    \dot{\strain}_{\vol} = 3\dot{\bar{\strain}}^{\ast} & \text{\revised{for the inelastic-strain-based method,}} \label{eq:vol_strain_eigen} \\[1em]
    \dot{\strain}_{\vol} = \dfrac{3\bar{K}^{\m}}{K+ \bar{K}^{\m}/\phi_{0}} \dot{\bar{\strain}}^{\ast} & \text{for the poromechanics-based method.} \label{eq:vol_strain_poro}
\end{subnumcases}
Notably, even if we keep the same amount of inelastic deformation $\dot{\bar{\strain}}^{\ast}$ generated by mineralization, the magnitudes of induced mechanical deformation are different across two approaches.
Interestingly, for the poromechanics-based approach, the volume change (\ie~$\dot{\strain}_{\vol}$) depends on the selections of $K$, $\bar{K}^{\m}$, and $\phi_{0}$.
For fixed $K$ and $\phi_{0}$, Eq.~\eqref{eq:vol_strain_poro} indicates that $\dot{\strain}_{\vol}$ increases monotonically with $\bar{K}^{\m}$, yet in a nonlinear asymptotic trend, with \revised{the minimum and maximum values of the poromechanics-based prediction} attained at two limits below,
\begin{subnumcases}{}
    \left( \dot{\strain}_{\vol}\right)_{\min} = 0  & \text{at $\bar{K}^{\m} = 0$,} \label{eq:vol_strain_min} \\[1em]
    \left( \dot{\strain}_{\vol}\right)_{\max} \rightarrow 3\phi_{0} \dot{\bar{\strain}}^{\ast}  & \text{when $\bar{K}^{\m} \gg K$. } \label{eq:vol_strain_max}
\end{subnumcases}
In this upper limit, the volumetric change reduces to a linear function of porosity, and it converges to the \revised{prediction of the inelastic-strain-based method} in Eq.~\eqref{eq:vol_strain_eigen} only if $\phi_{0} \rightarrow 1$, which is unattainable for real porous rocks.
Therefore, under this no-confinement condition, the induced volume change predicted by the poromechanics-based method is always smaller than that by the \revised{inelastic-strain-based} method.

\paragraph{Scenario II – Zero strain condition}
In this case, we have
\begin{align}
    \dot{\strain}_{ij} = 0. 
\end{align}
Therefore, we can obtain the corresponding total stress for both methods as follows based on Eqs.~\eqref{eq:stress_eigen_simplified} and \eqref{eq:stress_poro_simplified},
\begin{subnumcases}{}
    \dot{\stress}^\mathrm{inel}_{ij} = - 3K \dot{\bar{\strain}}^{\ast} \delta_{ij} & \text{\revised{for the inelastic-strain-based method,}} \label{eq:stress_eigen_confined} \\[1em]
    \dot{\stress}_{ij}^\mathrm{poro} =  - 3\bar{K}^{\m}\dot{\bar{\strain}}^{\ast} \delta_{ij}  & \text{for the poromechanics-based method.} \label{eq:stress_poro_confined}
\end{subnumcases}
It is clear that, under this full confinement condition, the difference in the induced total stress between two approaches lies in the choices of $K$ and $\bar{K}^{\m}$, and the two approaches yield the same results of the total stress if $\bar{K}^{\m}=K$.

\section{Discretization and solution algorithm}
\label{sec:numerical}


This section describes the numerical implementation of the chemo-hydro-mechanical system in GEOS~\citep{settgast2024geos}.
The fluid and species mass balance equations (Eqs.~\eqref{eq:flow-mass-balance-modified} and \eqref{eq:species-mass-balance-modified}) are discretized by the finite volume method (FVM), while the momentum balance (Eq.~\eqref{eq:momentum-balance}) is solved by the finite element method (FEM), with the two coupling approaches from Section~\ref{sec:coupling}, namely, the \revised{inelastic-strain-based} and poromechanics-based formulations, entering as constitutive relations at the material point.
\revised{Note that the inelastic-strain-based formulation implemented in GEOS follows \citet{evans2018poroelastic,evans2020phase}, as given in Eqs.~\eqref{eq:eigen-strain-evans} and \eqref{eq:eigen-strain-total-stress-evans}.}

\subsection{Coupled finite element–finite volume discretization}

To begin, we define a mesh $\mathcal{T}$ formed by nonoverlapping cells $E_i$ such that $\Omega \approx \Omega^{h} := \bigcup_i E_i$, where $(\cdot)^h$ denotes a discretized quantity, with each interior face $f$ shared by two neighboring cells.
Following the mixed FE--FV strategy for coupled flow-geomechanics simulation~\citep{white2019twostage,settgast2024geos}, the displacement $u_i$ is discretized using continuous, node-based shape functions, while the pore pressure $p_f$ and the species concentration $\ln C_{s}$ are discretized as cell-centered, piecewise-constant values.
Without loss of generality, we restrict our presentation to homogeneous prescribed displacement, \ie~$\bar{u}_i = 0$ on $\pd_u \Omega$.
Letting $\bar{\Omega} := \Omega \cup \pd \Omega$, the corresponding discrete trial spaces are then defined as,
\begin{align}
    \mathcal{V}_{u_i}^h &:= \left\{ \eta_i^h \;\middle|\; \eta_i^h \in C^0(\bar{\Omega}), \,\, \eta_i^h = 0 \text{ on } \pd_u \Omega, \,\, \eta_i^h|_{E} \in \mathbb{Q}_1(E) \,\, \forall E \in \mathcal{T} \right\}, \\
    \mathcal{V}_{p}^h &:= \left\{ \chi^h \;\middle|\; \chi^h \in L^2(\Omega), \,\, \chi^h|_{E} \in \mathbb{P}_0(E) \,\, \forall E \in \mathcal{T} \right\}, \\
    \mathcal{V}_{C}^h &:= \left\{ \psi^h \;\middle|\; \psi^h \in L^2(\Omega), \,\, \psi^h|_{E} \in \mathbb{P}_0(E) \,\, \forall E \in \mathcal{T} \right\},
\end{align}
with $i = 1, \ldots, n_\mathrm{dim}$, where $C^0(\bar{\Omega})$ and $L^2(\Omega)$ stand for the space of continuous functions on $\bar{\Omega}$ and square Lebesgue-integrable functions on $\Omega$, respectively, $\mathbb{Q}_1(E)$ is the space of multilinear polynomials mapped onto cell $E$, and $\mathbb{P}_0(E)$ is the piecewise-constant space on $E$.
This choice of cell-centered, piecewise-constant interpolation for $p_f$ and $C_s$ guarantees local (cell-wise) conservation of fluid mass and chemical species, which is an essential requirement for reactive transport problems.

Partitioning the time domain $\mathbb{T} = (0, t_{\max}]$ into discrete steps $t_i \in \left\{t_{0}, t_{1}, \ldots, t_{\max} \right \}$, with $\Delta t_{n+1} := t_{n+1} - t_{n}$ the time step size, the discrete residual equations are evaluated at $t_{n+1}$, given the converged solution at $t_n$.
For brevity, we hereafter omit the superscript for any quantity evaluated at the current time $t_{n+1}$, while a quantity evaluated at the previous, converged time level $t_n$ is denoted with the superscript $(\cdot)^n$.
The discrete residual equation for the momentum balance~\eqref{eq:momentum-balance} with zero body force then reads
\begin{align}
    \mathcal{R}^h_{u,i} = \int_{\Omega^h} \dfrac{\pd \eta_i^h}{\pd x_j} \stress_{ij} \, \dd V - \int_{\pd_t \Omega^h} \eta_i^h \, \bar{t}_{i} \, \dd A = 0 , \label{eq:residual-u}
\end{align}
where $\stress_{ij}$ is given by either Eq.~\eqref{eq:eigen-strain-total-stress-evans} or Eq.~\eqref{eq:poromech-final-stress}, depending on the coupling method adopted.
For the fluid mass balance~\eqref{eq:flow-mass-balance-modified},the discrete residual equation is written as
\begin{align}
    \mathcal{R}^h_{p} = \int_{\Omega^h} \chi^h \, \dfrac{\phi_\mathrm{eff} \rho_{f} - \phi_{\mathrm{eff}}^n \rho_{f}^n}{\Delta t} \, \dd V - \sum_{f \notin \pd_q \Omega} \jump{\chi^h}_f F^{f} - \int_{\Omega^h} \chi^h \, q_{f} \, \dd V = 0, \label{eq:residual-p}
\end{align}
where $q_{f}$ denotes the fluid sink/source term, $\jump{\cdot}_f$ is the jump of a quantity across an interior face $f$, and $F^f$ is the intercell fluid mass flux.
Likewise, the discrete residual equation for the species mass balance~\eqref{eq:species-mass-balance-modified} reads
\begin{align}
    \mathcal{R}^h_{s} = \int_{\Omega^h} \psi^h \, \dfrac{\phi_\mathrm{eff} C^\mathrm{total}_{s} - \phi_{\mathrm{eff}}^n \left(C^{\mathrm{total}}_{s} \right)^n}{\Delta t} \, \dd V - \sum_{f} \jump{\psi^h}_f F^{f}_{s} + \int_{\Omega^h} \psi^h \sum^{n_\mathrm{react}}_{m} \nu_{s,m} R_{m} \, \dd V = 0, \label{eq:residual-s}
\end{align}
where $F^f_s$ is the intercell species mass flux.
Following the standard two-point flux approximation (TPFA), the fluid flux across an interior face $f$ shared by cells $M$ and $N$ is evaluated as
\begin{align}
    F^{f} = -\dfrac{\rho_f^\mathrm{upw}}{\mu_f^\mathrm{upw}} \, \Upsilon^{f}_{k_0} \left[ \left(p_{f,N} + \widehat{\rho}_f g z_N \right) - \left(p_{f,M} + \widehat{\rho}_f g z_M \right) \right],
\end{align}
where $(\cdot)^\mathrm{upw}$ denotes an upwinded quantity, $\Upsilon^{f}_{k_0}$ is the face transmissibility computed from the permeability $k_0$, and $\widehat{\rho}_f$ is the arithmetic average of the fluid density at cells $M$ and $N$.
An analogous two-point expression is adopted for the species flux, combining an upstream-weighted advective term and a diffusive term,
\begin{align}
    F^{f}_{s} = \left(C_{s}^\mathrm{total}\right)^\mathrm{upw} F^{f} - \phi_0^\mathrm{upw} \, \Upsilon^{f}_{D} \left(C_{s,N}^\mathrm{total} - C_{s,M}^\mathrm{total}\right),
\end{align}
where $\Upsilon^{f}_{D}$ is the face transmissibility computed from the diffusion/dispersion coefficient $\mathcal{D}$.
In both cases, the face transmissibility for the permeability $k_0$ and the diffusion/dispersion coefficient $\mathcal{D}$ can be computed by the following general form,
\begin{align}
    \Upsilon^{f}_{\kappa} := \dfrac{\Upsilon^{f,M}_{\kappa} \, \Upsilon^{f,N}_{\kappa}}{\Upsilon^{f,M}_{\kappa} + \Upsilon^{f,N}_{\kappa}}, \quad \text{with} \quad \Upsilon^{f,M}_{\kappa} := |f| \dfrac{\left(x_f - x_M\right) \cdot \kappa_M \cdot n_f}{\left(x_f - x_M\right) \cdot \left(x_f - x_M\right)},
\end{align}
and 
\begin{align}
    \kappa =
    \begin{cases}
        k_0 & \text{for permeability transmissibility } \Upsilon^{f}_{k_0}, \\
        \mathcal{D} & \text{for diffusion/dispersion transmissibility } \Upsilon^{f}_{D},
    \end{cases}
\end{align}
where $|f|$ is the face area, $x_f$ is a collocation point on $f$, $x_M$ is the centroid of cell $M$, $\kappa_M$ is the value of $\kappa$ in cell $M$, and $n_f$ is the unit normal vector to $f$.  For cell $N$, $\Upsilon^{f,N}_{\kappa}$ is defined analogously.

\subsection{Sequential solution scheme for chemo-hydro-mechanical problems}
To solve the coupled chemo-hydro-mechanical system, we employ an iterative sequential approach that decouples the reactive transport (Eqs.~\eqref{eq:residual-s} and \eqref{eq:residual-p}) and mechanics (Eq.~\eqref{eq:residual-u}) subproblems with the fixed-strain splitting scheme~\citep{kim2011stability}.
The sequential solution algorithm is outlined in Algorithm~\ref{alg:fixed-strain}, where $\tensor{\mathcal{R}}_{\text{rt}}$ and $\tensor{\mathcal{R}}_{\text{mech}}$ denote the normalized residuals for the reactive transport and mechanics subproblems, respectively, and $\epsilon_{\text{rt}}$, $\epsilon_{\text{mech}}$, and $\epsilon_{\text{outer}}$ are the corresponding convergence tolerance parameters.
The key assumption of the fixed-strain split is that the volumetric strain $\varepsilon_{\vol} = \trace(\boldsymbol{\varepsilon})$ or equivalently the volumetric effective stress $\sigma'_{\vol} = K \varepsilon_{\vol}$ remains constant during the reactive transport subproblem solve.
This assumption is enforced through a modified porosity evolution equation that accounts for both mechanical deformation and chemical reactions.
Following Eq.~\eqref{eq:isotropic-porosity-poromech-expanded}, the porosity at sequential iteration $k+1$ is given by:
\begin{align}
\phi^{(n+1,k+1)} = \phi^n + \frac{1}{1 + \frac{1}{\mathcal{N}}\frac{\bar{K}^{\m}}{\phi_0}} \Bigg[ &\alpha\dfrac{ \left(\sigma'_{\vol}\right)^{(n+1,k)} - \left(\sigma'_{\vol}\right)^{n}}{K} \nonumber \\
&+ \dfrac{1}{\mathcal{N}}\left(\frac{\bar{K}^{\m}+K^{f}}{K^{f}}(p_f^{(n+1,k+1)} - p_f^n) + 3\bar{K}^{\m} {\left(\Delta\bar{\strain}^\ast\right)}^n \right) \Bigg], \label{eq:porosity-fixed-strain}
\end{align}
where $\frac{1}{\mathcal{N}} = \frac{\alpha - \phi_0}{K^{\mathrm{s}}}$ is the inverse of the Biot tangent modulus.
For simplicity, we apply a semi-implicit update scheme where \revised{$\Delta\bar{\strain}^\ast$} uses reaction rates from the last converged timestep rather than the current outer iteration, thereby avoiding complex derivative calculations in the porosity update.
Despite being only conditionally stable, the scheme performs robustly for the problems in this study, with convergence typically achieved within five outer iterations per timestep.

\begin{algorithm}[t]
\caption{Fixed-strain split for chemo-hydro-mechanics.}
\label{alg:fixed-strain}
\begin{algorithmic}[1]
\Require $\tensor{u}^{n}$, $p_f^{n}$, $\ln C_s^{n}$, $\Delta t$
\Ensure $\tensor{u}$, $p_f$, $\ln C_s$ at time $t^{n+1} = t^n + \Delta t$
\State Initialize $\tensor{u} \leftarrow \tensor{u}^{n}$, $p_f \leftarrow p_f^{n}$, $\ln C_s \leftarrow \left (\ln C_s \right )^{n}$
\Repeat
    \State \textbf{Reactive transport subsolve:}
    \State Update porosity $\phi$ using Eq.~\eqref{eq:porosity-fixed-strain}
    \Repeat
        \State Assemble system with updated $\phi$
        \State Solve the linearized system of Eqs.~\eqref{eq:residual-p} and \eqref{eq:residual-s} for $\delta p_f$, $\delta \ln C_s$
        \State Update $p_f \leftarrow p_f + \delta p_f$, $\ln C_s \leftarrow \ln C_s + \delta \ln C_s$
    \Until{$\|\tensor{\mathcal{R}}_{\text{rt}}\| < \epsilon_{\text{rt}}$}
    \State Update $\bar{p}_{\m}$ via Eq.~\eqref{eq:isotropic-mineral-stress-poromech}
    \State \textbf{Mechanics subsolve:}
    \Repeat
        \State Assemble system with updated $p_f$ and $\bar{p}_{\m}$
        \State Solve the linearized form of Eq.~\eqref{eq:residual-u} for $\delta \tensor{u}$
        \State Update $\tensor{u} \leftarrow \tensor{u} + \delta \tensor{u}$
    \Until{$\|\tensor{\mathcal{R}}_{\text{mech}}\| < \epsilon_{\text{mech}}$}
    \State Evaluate $\mathcal{E} \leftarrow \sqrt{\|\tensor{\mathcal{R}}_{\text{rt}}\|^2 + \|\tensor{\mathcal{R}}_{\text{mech}}\|^2}$
\Until{$\mathcal{E} < \epsilon_{\text{outer}}$}
\State \Return $\tensor{u}$, $p_f$, $\ln C_s$
\end{algorithmic}
\end{algorithm}

\section{Numerical demonstration}
\label{sec:results}


This section presents numerical examples of coupled chemo-mechanical problems to demonstrate the capabilities of the proposed poromechanics-based coupling framework.
In selected examples, results from the poromechanics-based method are thoroughly compared against those from the \revised{inelastic-strain-based} method (following \citealp{evans2018poroelastic,evans2020phase}), highlighting the distinct mechanical predictions arising from the two approaches.
Advanced examples further demonstrate mineralization-induced cracking in various geological contexts.
In particular, the second and third examples are motivated by the two reaction-induced fracturing mechanisms~\citep{xing2018generating}: (i) the second example studies the stress and cracking driven by differential volume expansion across confined regions with different reactivity~\citep{zhu2016experimental,xing2018generating,zheng2019mixed}; (ii) the third example captures fracturing caused directly by the mineralization (crystallization) force exerted by pore-scale mineral growth~\citep{weyl1959pressure,scherer2004stress,kelemen2012reaction}.
Through these examples, we also explore the applicability of both poromechanics-based and \revised{inelastic-strain-based} methods to each fracturing mechanism.

\subsection{Serpentinization with 1D surface area gradient}
\label{sec:1d-serpentinization}
\begin{figure}[htbp]
    \centering
    \includegraphics[width=\textwidth]{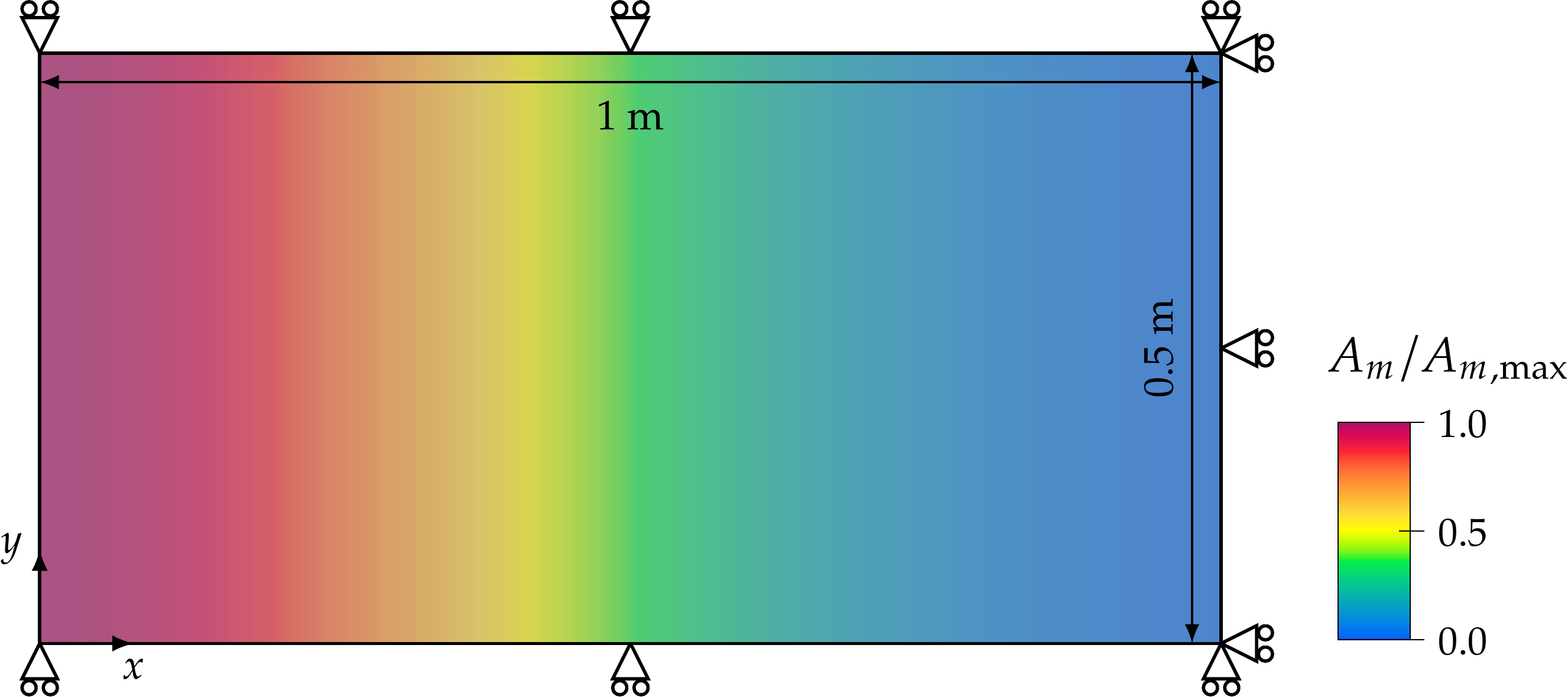}
    \caption{Serpentinization with 1D surface area gradient: problem setup.}
    \label{fig:1d-serpentinization-setup}
\end{figure}

We begin with a simple problem in which the mineral surface areas vary spatially along one direction across the domain.
As illustrated in Figure~\ref{fig:1d-serpentinization-setup}, the problem domain is a rectangle with 1~m in length and 0.5~m in width.
For the mechanical boundary conditions, the top, bottom, and right boundaries are constrained by rollers, while the left boundary is traction-free, allowing the domain to expand freely in the horizontal direction.
The pore fluid pressure is kept constant at zero throughout the domain (\ie~$p_f = 0$), such that no fluid flux occurs and the chemo-mechanical coupling is driven solely by mineral reactions.

This example adopts a geochemical system based on serpentinization, which involves the following mineral reactions:
\begin{align}
    \ce{Mg2SiO4 (forsterite) + 4H+ &<=> 2Mg^2+ + SiO2(aq) + 2H2O}, \label{eq:forsterite} \\
    \ce{Mg3Si2O5(OH)4 (serpentine) + 6H+ &<=> 3Mg^2+ + 2SiO2(aq) + 5H2O}, \label{eq:serpentine} \\
    \ce{Mg(OH)2 (brucite) + 2H+ &<=> Mg^2+ + 2H2O}. \label{eq:brucite}
\end{align}
The kinetic rate constants and equilibrium constants for these reactions are listed in Table~\ref{tab:reaction-properties}.
All three minerals, namely, forsterite, serpentine, and brucite, share the same exponential profile below for their \emph{normalized} surface areas ($A_m/A_{m,\max}$), while the maximum surface areas $A_{m,\max}$ differ across three minerals (see Table~\ref{tab:1d-serpentinization-properties}),
\begin{align}
    A_m/A_{m, \max} = \exp\left[-(x/0.5)^2\right] .
\end{align}
The essential material properties and initial species concentrations used in this example are also summarized in Table~\ref{tab:1d-serpentinization-properties}.
Notably, the initial porosity is set to $\phi_0 = 1.0$, which is not physically realistic but serves to better compare the two coupling schemes under an extreme condition, as motivated by the analytical comparison in Section~\ref{sec:analytical-comparison}.
Additionally, to further simplify the comparison, the porosity and mineral surface areas are held fixed at their initial values throughout the simulation.
\begin{table}[htbp]
\caption{Kinetic and thermodynamic parameters for the serpentinization reactions.}
\label{tab:reaction-properties}
\centering
\begin{tabular}{p{4.0cm}ccccc}
\toprule
\textbf{Property} & \textbf{Symbol} & \textbf{Forsterite} & \textbf{Serpentine} & \textbf{Brucite} & \textbf{Unit} \\
\midrule
Rate constant      & $k_m$               & $2.29\times10^{-11}$ & $10^{-12}$ & $5.75\times10^{-9}$ & mol\,m$^{-2}$\,s$^{-1}$ \\
Equilibrium constant & $K^\mathrm{eq}_{m}$ & $1.4\times10^{28}$ & $3.54\times10^{31}$ & $2.75\times10^{16}$ & {--}                    \\
\bottomrule
\end{tabular}
\end{table}

\begin{table}[htbp]
\caption{Serpentinization with 1D surface area gradient: material properties and initial conditions. Properties marked with $(\dagger)$ are specific to the poromechanics-based coupling model.}
\label{tab:1d-serpentinization-properties}
\centering
\begin{tabular}{p{2.3cm}p{5.0cm}ccc}
\toprule
\multicolumn{1}{l}{\textbf{Category}} &
\multicolumn{1}{l}{\textbf{Property}} &
\textbf{Symbol} & \textbf{Value} &
\multicolumn{1}{c}{\textbf{Unit}} \\
\midrule
\multirow{4}{*}{\parbox{2.3cm}{Rock matrix}}
 & Drained bulk modulus        & $K$         & 6.67 & GPa \\
 & Shear modulus               & $G$         & 4.00 & GPa \\
 & Initial porosity          & $\phi_0$    & 1.0 & --  \\
 & Solid grain bulk modulus$^{\dagger}$ & $K^{\mathrm{s}}$ & $10^{20}$ & GPa \\
\midrule
\multirow{16}{*}{\parbox{2.3cm}{Minerals}}
 & Mineral bulk modulus$^{\dagger}$ & $\bar{K}^{\m}$            & $K,\ 10K,\ 100K$ & GPa      \\
\cmidrule{2-5}
 & \multicolumn{4}{l}{\textit{Forsterite (Fo)}} \\
 & \quad Initial volume fraction    & $\theta_{\mathrm{Fo},0}$    & 0.5 & --         \\
 & \quad Molar weight               & $M_{\mathrm{Fo}}$         & 140.69 & g/mol      \\
 & \quad Mineral density            & $\rho_{\mathrm{Fo}}$      & 3212.9 & kg/m$^3$   \\
 & \quad Maximum surface area       & $A_{\mathrm{Fo},\max}$    & $1.3\times10^{4}$  & m$^2$/m$^3$ \\
\cmidrule{2-5}
 & \multicolumn{4}{l}{\textit{Serpentine (Serp)}} \\
 & \quad Initial volume fraction    & $\theta_{\mathrm{Serp},0}$   & 0.0 & --         \\
 & \quad Molar weight               & $M_{\mathrm{Serp}}$        & 277.11 & g/mol      \\
 & \quad Mineral density            & $\rho_{\mathrm{Serp}}$     & 2554.03 & kg/m$^3$   \\
 & \quad Maximum surface area       & $A_{\mathrm{Serp},\max}$   & $2\times10^{5}$ & m$^2$/m$^3$ \\
\cmidrule{2-5}
 & \multicolumn{4}{l}{\textit{Brucite (Bruc)}} \\
 & \quad Initial volume fraction    & $\theta_{\mathrm{Bruc},0}$   & 0.0 & --         \\
 & \quad Molar weight               & $M_{\mathrm{Bruc}}$        & 58.32 & g/mol      \\
 & \quad Mineral density            & $\rho_{\mathrm{Bruc}}$     & 2367.84 & kg/m$^3$   \\
 & \quad Maximum surface area       & $A_{\mathrm{Bruc},\max}$   & 50 & m$^2$/m$^3$ \\
\midrule
\multirow{3}{*}{\parbox{2.3cm}{Initial\\concentrations}}
 & Hydrogen ion, \ce{H+}        & $C_{\mathrm{H^+},0}$         & $5.62\times10^{-10}$  & mol/kg \\
 & Magnesium ion, \ce{Mg^{2+}}  & $C_{\mathrm{Mg^{2+}},0}$    & $10^{-2}$ & mol/kg \\
 & Aqueous silica, \ce{SiO2(aq)} & $C_{\mathrm{SiO_2},0}$     & $10^{-5}$ & mol/kg \\
\bottomrule
\end{tabular}
\end{table}

\begin{figure}[htbp]
    \centering
    \subfloat[]{\includegraphics[width=0.48\textwidth]{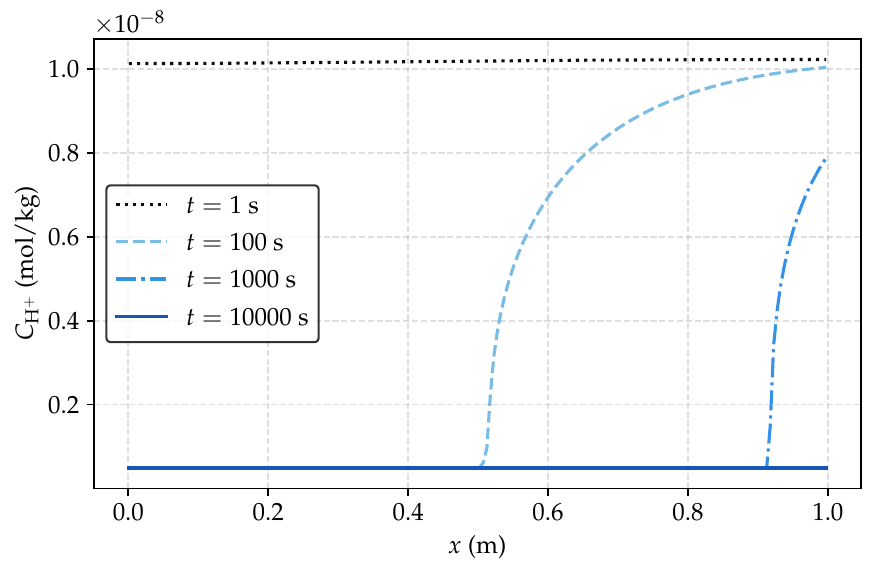}\label{fig:1d-serpentinization-concH}}
    \hfill
    \subfloat[]{\includegraphics[width=0.48\textwidth]{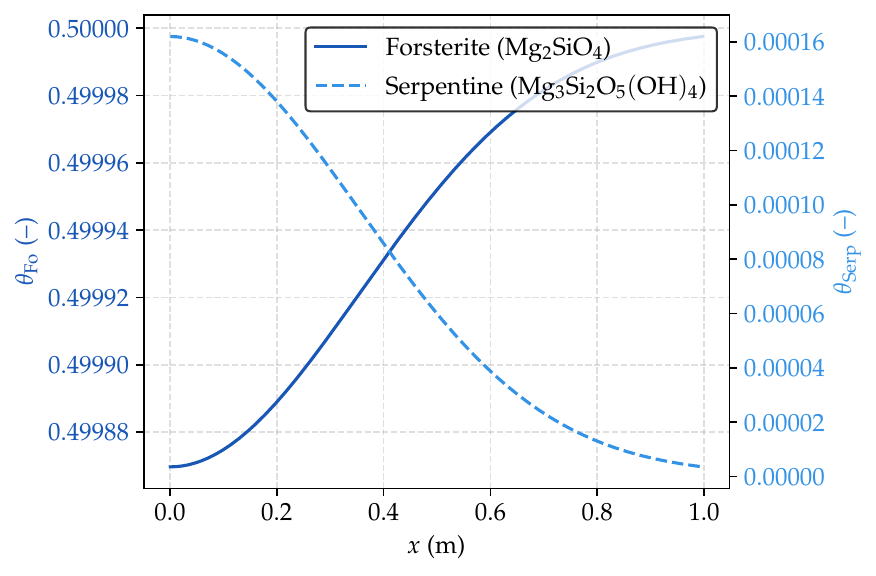}\label{fig:1d-serpentinization-volFrac}}
    \caption{Serpentinization with 1D surface area gradient: (a) spatio-temporal evolution of $\ce{H+}$ concentration; (b) spatial profile of volume fractions for forsterite and serpentine. }
    \label{fig:1d-serpentinization-reaction-results}
\end{figure}

Figures~\ref{fig:1d-serpentinization-concH} and \ref{fig:1d-serpentinization-volFrac} present the evolution of the $\ce{H+}$ concentration and the mineral volume fractions for forsterite and serpentine, respectively.
Since the porosity and mineral surface areas are held constant at identical initial values in both \revised{inelastic-strain-based} and poromechanics-based methods, the chemical reaction results including species concentrations and mineral volume fraction changes are the same between the two coupling approaches.
As the serpentinization reaction proceeds, the $\ce{H+}$ concentration decreases monotonically with time, reflecting the overall consumption of hydrogen ions by the mineral reactions.
The volume fraction profile shows that forsterite dissolves across the domain while serpentine precipitates. 
Brucite is not shown here as its volume fraction change is negligible due to the small initial surface area.
The net effect is a positive change in the total mineral volume fraction, indicating an overall volumetric expansion of the system.
Spatially, both $\ce{H+}$ depletion and mineral volume fraction changes are more pronounced on the left side of the domain, where the surface areas are higher, and decrease toward the right where the surface areas approach zero.

\begin{figure}[htbp]
    \centering
    \subfloat[\revised{Poromechanics-based}]{\includegraphics[width=0.48\textwidth]{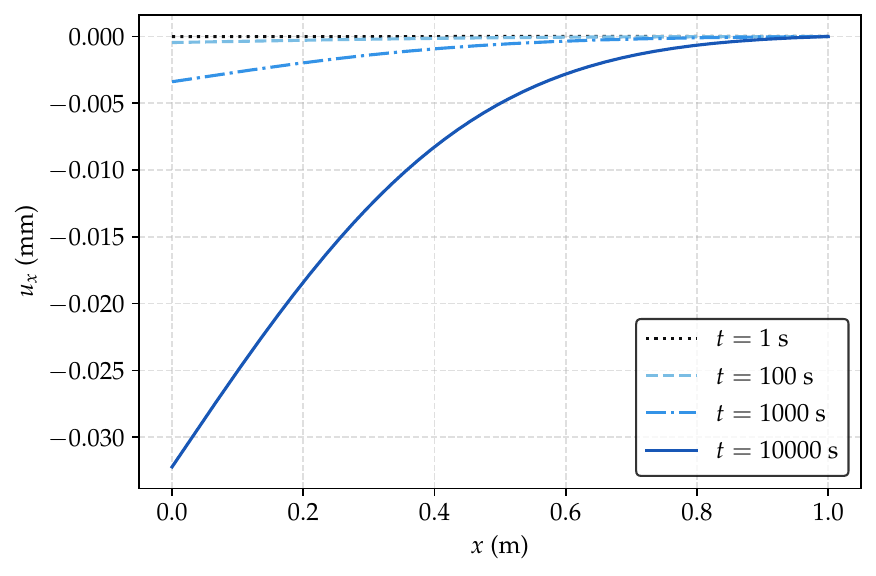}\label{fig:1d-serpentinization-poromech-dispX}}
    \hfill
    \subfloat[\revised{Inelastic-strain-based}]{\includegraphics[width=0.48\textwidth]{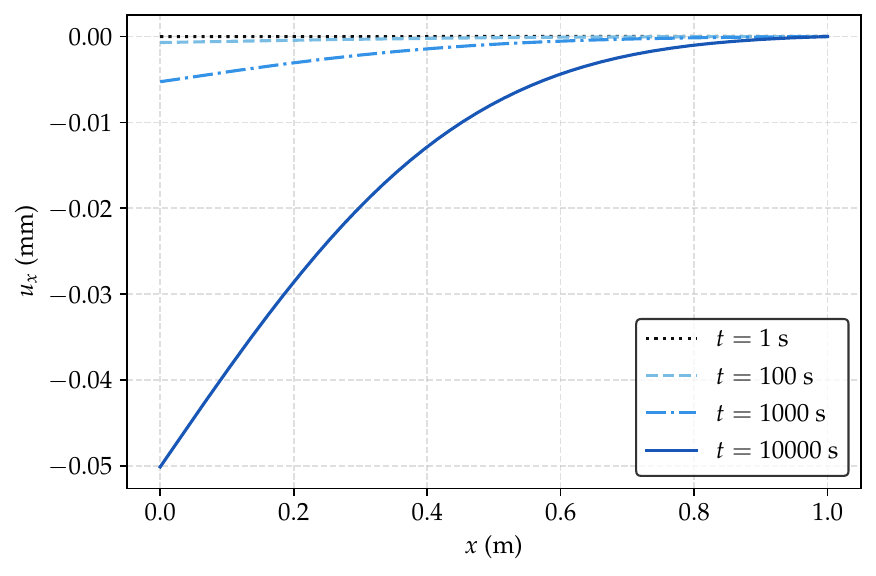}\label{fig:1d-serpentinization-eigenStrain-dispX}}
    \caption{Serpentinization with 1D surface area gradient: \revised{spatio-temporal evolution of $u_{x}$ predicted by the poromechanics-based and inelastic-strain-based methods.}}
    \label{fig:1d-serpentinization-compare-dispX}
\end{figure}

\begin{figure}[htbp]
    \centering
    \subfloat[\revised{Poromechanics-based}]{\includegraphics[width=0.48\textwidth]{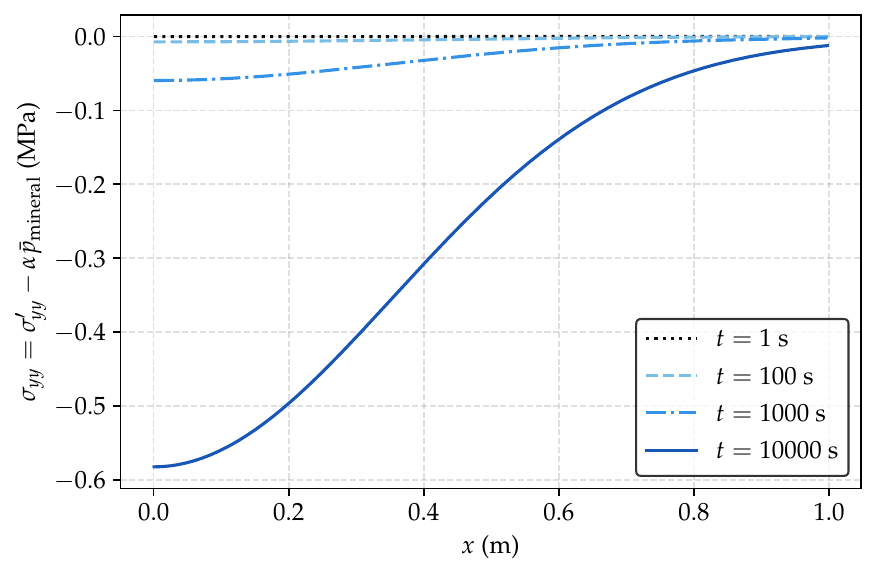}\label{fig:1d-serpentinization-poromech-totalStressYY}}
    \hfill
    \subfloat[\revised{Inelastic-strain-based}]{\includegraphics[width=0.48\textwidth]{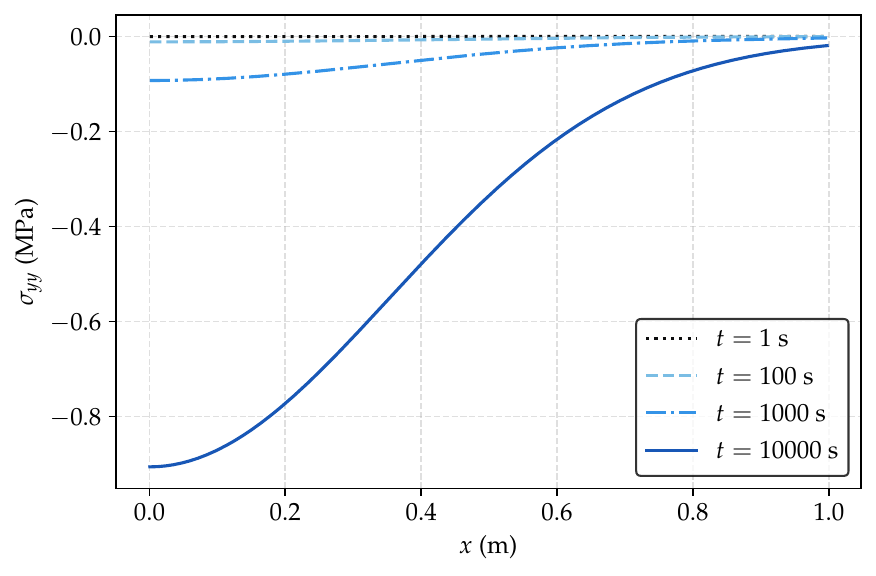}\label{fig:1d-serpentinization-eigenStrain-totalStressYY}}
    \caption{Serpentinization with 1D surface area gradient: \revised{spatio-temporal evolution of $\sigma_{yy}$ predicted by the poromechanics-based and inelastic-strain-based methods.}}
    \label{fig:1d-serpentinization-compare-totalStressYY}
\end{figure}

Next, we compare the reaction-induced volumetric change and the total stress predicted by both chemo-mechanics coupling methods in Figures~\ref{fig:1d-serpentinization-compare-dispX} and \ref{fig:1d-serpentinization-compare-totalStressYY}, respectively, with $\bar{K}^{\m} = K$ adopted for the poromechanics-based method.
Consistent with the net positive change in mineral volume fraction, both methods correctly predict volumetric expansion of the domain, as reflected in the negative horizontal displacement $u_x$ (the domain expands toward the traction-free left boundary).
Meanwhile, both methods predict compressive total stress in the $y$ direction, as the rollers on the top and bottom boundaries constrain volumetric expansion.
When comparing the induced displacement, the \revised{inelastic-strain-based} method predicts larger magnitude of $u_x$ than the poromechanics-based method.
This agrees well with the analytical result under the zero-stress condition (Eqs.~\eqref{eq:vol_strain_eigen} and \eqref{eq:vol_strain_poro}), which indicates that the poromechanics-based method always predicts smaller volumetric expansion than the \revised{inelastic-strain-based} method for the no confinement case unless $\bar{K}^{\m} \gg  K $ and $ \phi \to 1$.
Furthermore, the magnitude of the total stress predicted by the poromechanics-based method is smaller than that of the \revised{inelastic-strain-based} method, which may seem to contradict the zero-strain analytical result in Eqs.~\eqref{eq:stress_eigen_confined} and \eqref{eq:stress_poro_confined}, where the two methods yield identical stress when $\bar{K}^{\m} = K$.
The key distinction is that the zero-strain result applies strictly under full confinement, whereas this example employs mixed boundary conditions with a traction-free left boundary that permits partial expansion.
Since the poromechanics-based method has inherently less expansion, the domain reaches mechanical equilibrium with a smaller strain in the constrained direction, which in turn results in lower stress build-up compared to the \revised{inelastic-strain-based} method.

\begin{figure}[htbp]
    \centering
    \subfloat[]{\includegraphics[width=0.48\textwidth]{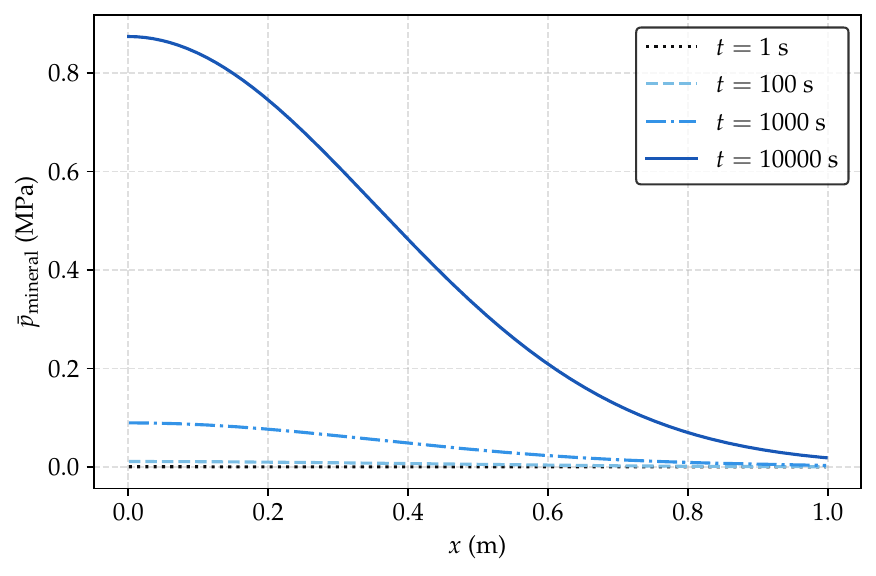}\label{fig:1d-serpentinization-poromech-mineralPres}}
    \hfill
    \subfloat[]{\includegraphics[width=0.48\textwidth]{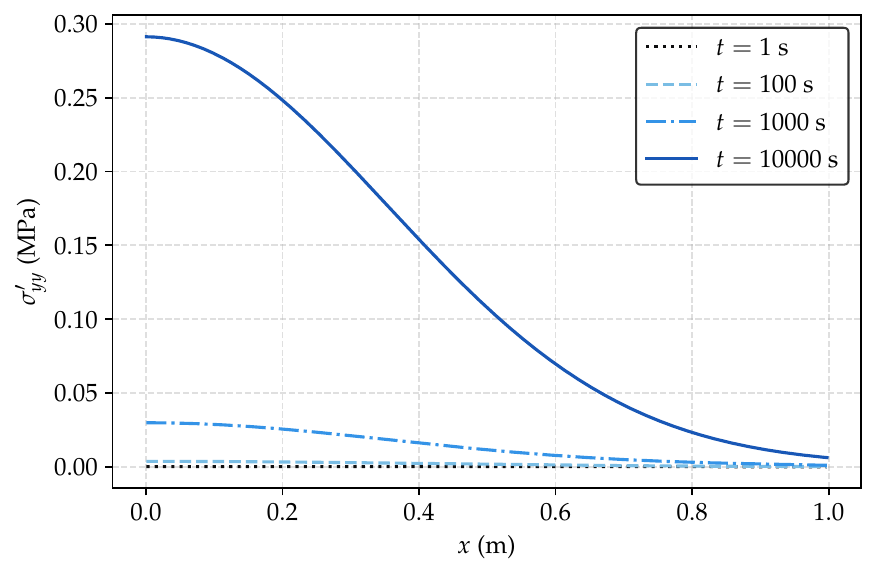}\label{fig:1d-serpentinization-poromech-effStressYY}}
    \caption{Serpentinization with 1D surface area gradient: (a) spatio-temporal evolution of mineral pressure $\bar{p}_{\m}$ by the poromechanics-based method; (b) spatio-temporal evolution of effective stress $\sigma'_{yy}$ by the poromechanics-based method. }
    \label{fig:1d-serpentinization-poromech-results}
\end{figure}

A unique advantage of the poromechanics-based method over the \revised{inelastic-strain-based} method is its ability to separately resolve the pore mineral pressure $\bar{p}_{\m}$ and the effective stress $\stress'_{ij}$ in the host rock, whereas the \revised{inelastic-strain-based} method only provides a single lump-sum stress for the solid phase.
Figures~\ref{fig:1d-serpentinization-poromech-mineralPres} and \ref{fig:1d-serpentinization-poromech-effStressYY} present the spatio-temporal evolutions of $\bar{p}_{\m}$ and the effective stress $\sigma'_{yy}$, respectively, as predicted by the poromechanics-based method.
As the serpentinization reaction proceeds, the mineral pressure builds up with time due to the net growth of minerals in the pore space, reaching its maximum on the left side of the domain where the reaction rate is the highest.
Notably, while the total stress $\sigma_{yy}$ is compressive throughout the domain as shown in Figure~\ref{fig:1d-serpentinization-poromech-totalStressYY}, the induced effective stress $\sigma'_{yy}$ is tensile, indicating the potential for Mode~I cracking in the host rock driven by pore expansion due to mineralization.
This physically important distinction between total stress and effective stress cannot be captured by the \revised{inelastic-strain-based} method, which provides only a single stress term without decomposing the contributions from the rock skeleton and the pore minerals.

\begin{figure}[htbp]
    \centering
    \subfloat[]{\includegraphics[width=0.48\textwidth]{Figures/1DVaryingSurfaceArea_Poromech_mineralBulkMod=1bulkMod_dispX.pdf}\label{fig:1d-serpentinization-poromech-dispX-1K}}
    \hfill
    \subfloat[]{\includegraphics[width=0.48\textwidth]{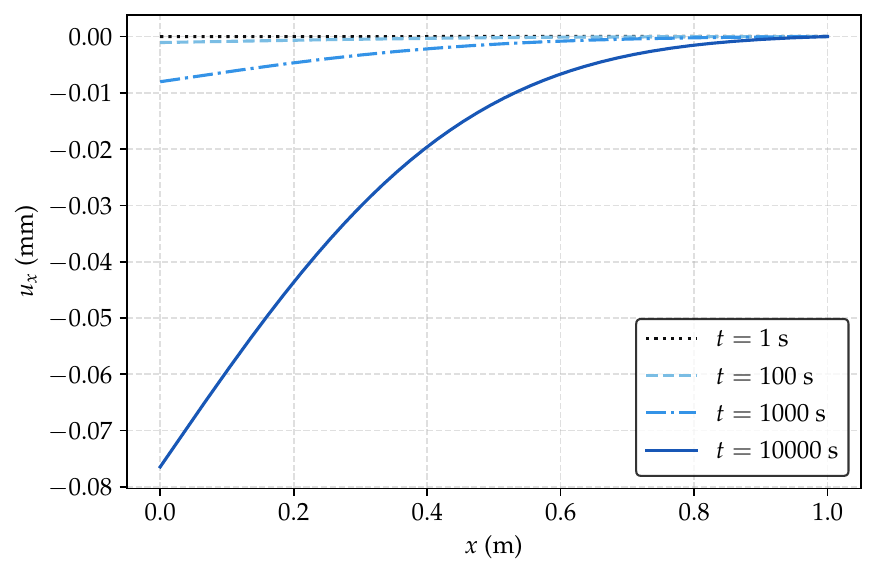}\label{fig:1d-serpentinization-poromech-dispX-10K}}
    \\
    \makebox[\textwidth][c]{\subfloat[]{\includegraphics[width=0.48\textwidth]{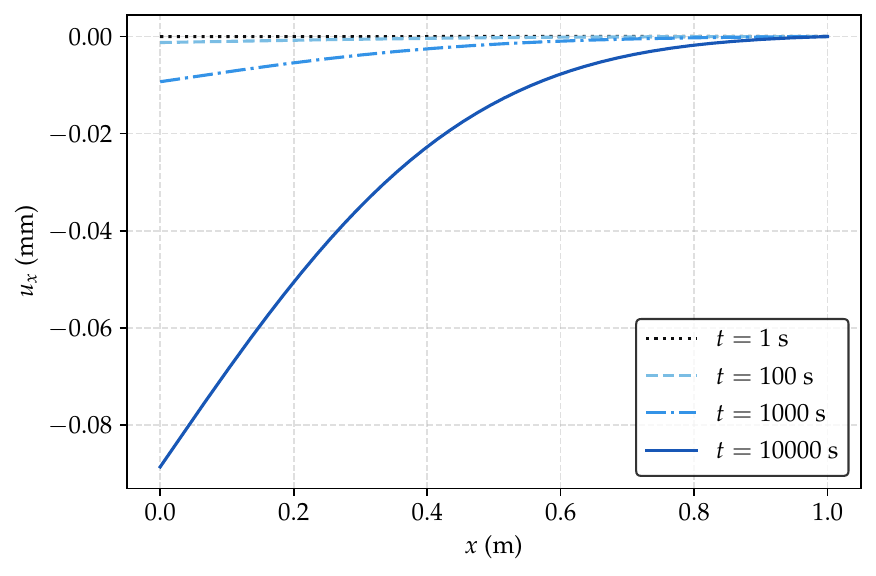}\label{fig:1d-serpentinization-poromech-dispX-100K}}}
    \caption{Serpentinization with 1D surface area gradient: spatio-temporal evolution of $u_x$ by the poromechanics-based method with (a) $\bar{K}^{\m} = K$, (b) $\bar{K}^{\m} = 10K$, and (c) $\bar{K}^{\m} = 100K$.}
    \label{fig:1d-serpentinization-poromech-dispX-Km}
\end{figure}

Additionally, three values of the mineral bulk modulus are considered for the poromechanics-based model, namely $\bar{K}^{\m} = K$, $10K$, and $100K$, to assess the sensitivity of the mechanical response to this parameter.
As shown in Figure~\ref{fig:1d-serpentinization-poromech-dispX-Km}, the magnitude of the induced displacement $u_x$ increases with $\bar{K}^{\m}$, but with a diminishing rate, indicating a nonlinear relationship.
This is consistent with the analytical result under the zero-stress condition in Eq.~\eqref{eq:vol_strain_poro}, where the volumetric expansion is proportional to $1/(K/\bar{K}^{\m} + 1/\phi_{0})$, a nonlinear and monotonically increasing function of $\bar{K}^{\m}$, whose rate of increase also diminishes as $\bar{K}^{\m}$ grows larger.

\subsection{Serpentinization induced radial core expansion and cracking}

\begin{figure}[htbp]
    \centering
    \includegraphics[height=0.4\textheight]{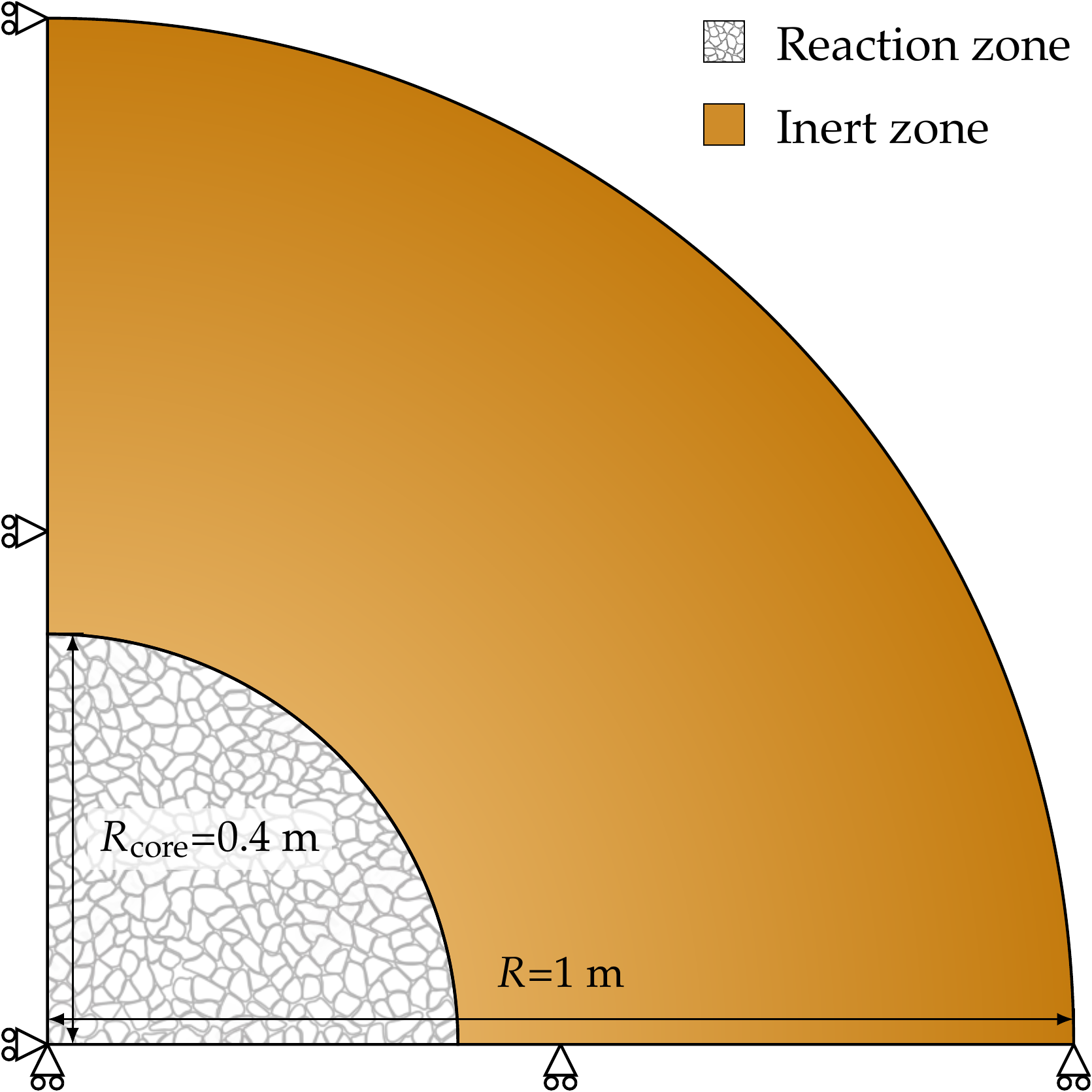}
    \caption{Serpentinization induced radial core expansion and cracking: problem setup.}
    \label{fig:quarter-cylinder-setup}
\end{figure}

The second example examines the ability of the poromechanics-based method to capture mineralization-induced tensile stress and potential cracking associated with differential volume expansion in a more realistic geometry.
This example also adopts the same serpentinization reactions (Eqs.~\eqref{eq:forsterite}--\eqref{eq:brucite}).
As depicted in Figure~\ref{fig:quarter-cylinder-setup}, the problem domain is a quarter-cylinder, taking advantage of the symmetry to reduce computational cost.
The inner core region is assigned nonzero surface areas for all three minerals, allowing the serpentinization reaction to proceed, while the outer annulus is treated as an inert zone with zero surface areas.
For the mechanical boundary conditions, rollers are applied along the two straight edges (left and bottom) to enforce the symmetry conditions, while the outer circumference is traction-free.
This setup is inspired by laboratory experiments that investigated volume expansion and induced fracturing due to carbonation or hydration reactions~\citep{zhu2016experimental,xing2018generating,zheng2019mixed}, in which a highly porous and reactive sample (\eg~olivine sand) is packed inside a hollow cylindrical cup made of a tight, low-permeability rock.
As the reactive core expands due to mineral precipitation, it induces tensile hoop stress in the surrounding outer rock that may eventually drive fractures therein and enhance the permeability and reactive transport.

The material properties and initial species concentrations for this example are summarized in Table~\ref{tab:quarter-cylinder-properties}, with mineral molar weights and densities identical to those in the first example (see Table~\ref{tab:1d-serpentinization-properties}) and omitted here for brevity.
Since mineral reactions only occur in the inner core, the parameters associated with chemo-mechanical coupling (\ie~$K^{\mathrm{s}}$, $\bar{K}^{\m}$, and the initial mineral volume fractions) are only specified for the reaction region.

\begin{table}[htbp]
\caption{Serpentinization induced radial core expansion and cracking: material properties and initial conditions at the reaction region (inner core) and the inert region (outer annulus). Properties marked with $(\dagger)$ are specific to the poromechanics-based coupling model.}
\label{tab:quarter-cylinder-properties}
\centering
\begin{tabular}{p{2.3cm}p{5.0cm}ccc c}
\toprule
\multicolumn{1}{l}{\textbf{Category}} &
\multicolumn{1}{l}{\textbf{Property}} &
\textbf{Symbol} & \textbf{Reaction zone} & \textbf{Inert zone} &
\multicolumn{1}{c}{\textbf{Unit}} \\
\midrule
\multirow{6}{*}{\parbox{2.3cm}{Rock matrix}}
 & Drained bulk modulus        & $K$         & 6.67 & 33.33 & GPa \\
 & Shear modulus               & $G$         & 4.00 & 20.00 & GPa \\
 & Initial porosity            & $\phi_0$    & 0.1 & 0.01 & --  \\
 & Solid grain bulk modulus$^{\dagger}$ & $K^{\mathrm{s}}$ & $10^{20}$ & --- & GPa \\
 & Reference permeability        & $k_0$       & \multicolumn{2}{c}{$3\times10^{-18}$} & m$^2$ \\
 & Reference diffusivity       & $\mathcal{D}_{0}$ & \multicolumn{2}{c}{$10^{-9}$} & m$^2$/s \\
\midrule
\multirow{2}{*}{\parbox{2.3cm}{Fluid}}
 & Dynamic viscosity    & $\mu_f$    & \multicolumn{2}{c}{$10^{-3}$} & Pa$\cdot$s \\
 & Compressibility      & $c_f$      & \multicolumn{2}{c}{$5\times10^{-10}$} & Pa$^{-1}$ \\
\midrule
\multirow{10}{*}{\parbox{2.3cm}{Minerals}}
 & Mineral bulk modulus$^{\dagger}$ & $\bar{K}^{\m}$ & 66.67 & --- & GPa \\
\cmidrule{2-6}
 & \multicolumn{5}{l}{\textit{Forsterite (Fo)}} \\
 & \quad Initial volume fraction    & $\theta_{\mathrm{Fo},0}$    & 0.5 & --- & --         \\
 & \quad Initial surface area       & $A_{\mathrm{Fo},0}$    & $1.3\times10^{5}$ & 0 & m$^2$/m$^3$ \\
\cmidrule{2-6}
 & \multicolumn{5}{l}{\textit{Serpentine (Serp)}} \\
 & \quad Initial volume fraction    & $\theta_{\mathrm{Serp},0}$  & 0.0 & --- & --         \\
 & \quad Initial surface area       & $A_{\mathrm{Serp},0}$  & $2\times10^{6}$ & 0 & m$^2$/m$^3$ \\
\cmidrule{2-6}
 & \multicolumn{5}{l}{\textit{Brucite (Bruc)}} \\
 & \quad Initial volume fraction    & $\theta_{\mathrm{Bruc},0}$  & 0.0 & --- & --         \\
 & \quad Initial surface area       & $A_{\mathrm{Bruc},0}$  & 500 & 0 & m$^2$/m$^3$ \\
\midrule
\multirow{3}{*}{\parbox{2.3cm}{Initial\\concentrations}}
 & Hydrogen ion, \ce{H+}         & $C_{\mathrm{H^+},0}$      & $5.62\times10^{-10}$ & $10^{-7}$ & mol/kg \\
 & Magnesium ion, \ce{Mg^{2+}}   & $C_{\mathrm{Mg^{2+}},0}$ & $10^{-2}$ & $10^{-6}$ & mol/kg \\
 & Aqueous silica, \ce{SiO2(aq)} & $C_{\mathrm{SiO_2},0}$   & $10^{-2}$ & $10^{-9}$ & mol/kg \\
\bottomrule
\end{tabular}
\end{table}

\begin{figure}[htbp]
    \centering
    \includegraphics[width=\textwidth]{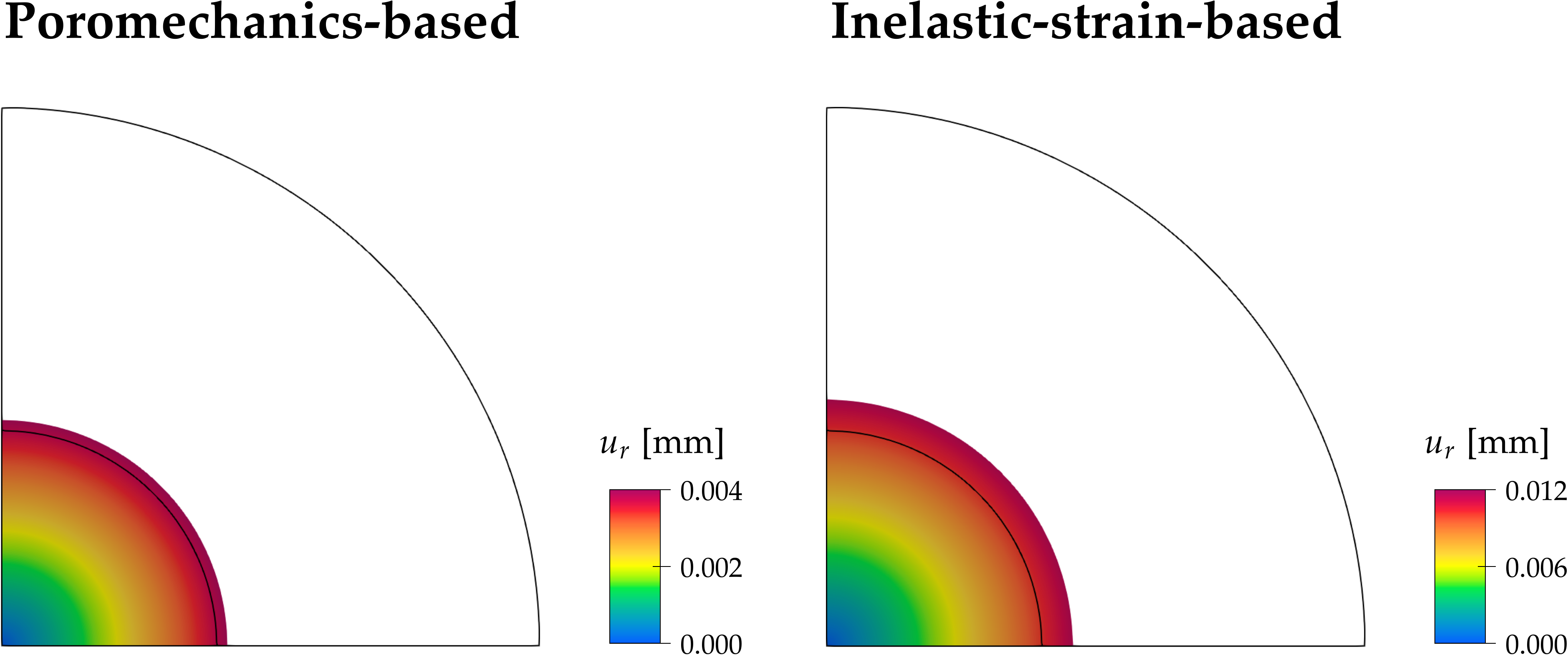}
    \caption{Serpentinization induced radial core expansion: comparison of the induced radial displacement predicted by the poromechanics-based and \revised{inelastic-strain-based} methods at $t=1000$ s. Deformation is magnified by a factor of 5000.}
    \label{fig:quarter-cylinder-compare-dispR}
\end{figure}

\begin{figure}[htbp]
    \centering
    \includegraphics[width=\textwidth]{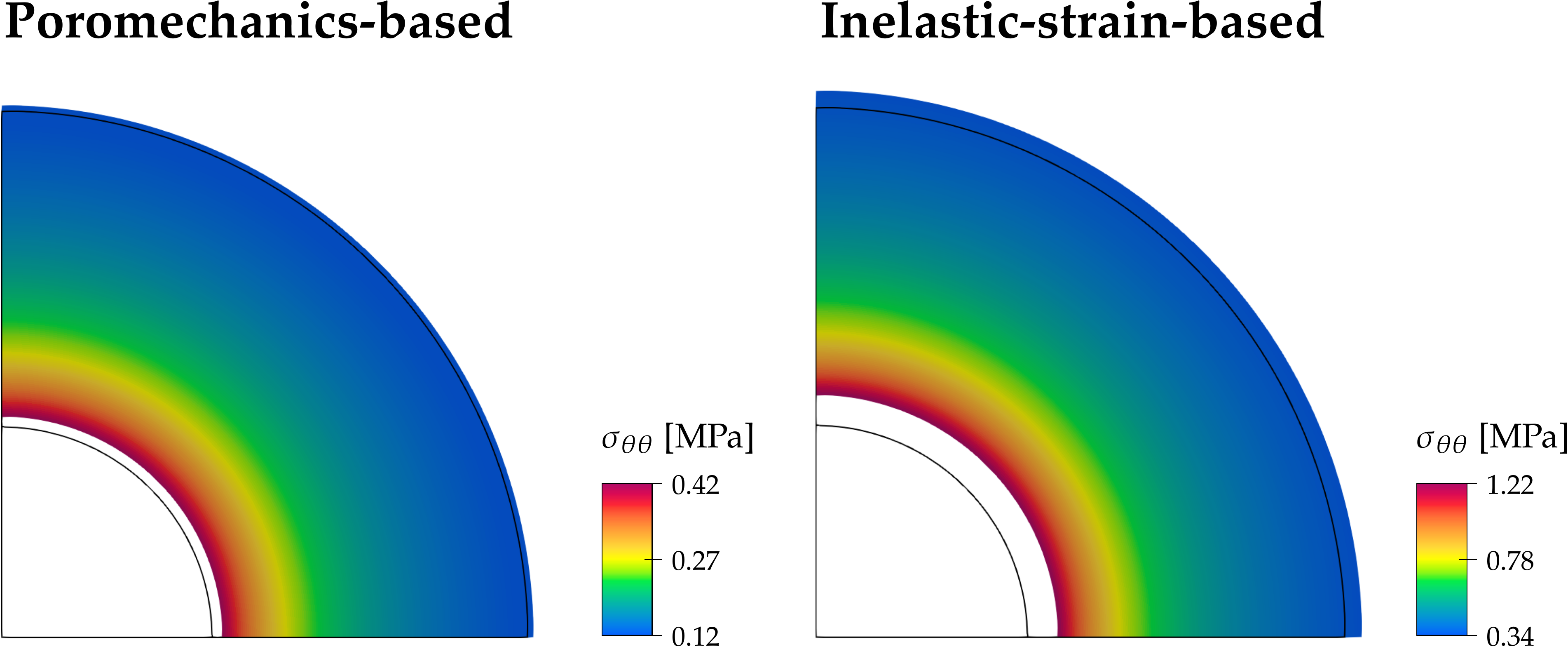}
    \caption{Serpentinization induced radial core expansion and cracking: comparison between the poromechanics-based and \revised{inelastic-strain-based} methods in induced angular hoop stress of the inert zone at $t=1000$ s. Deformation is magnified by a factor of 5000.}
    \label{fig:quarter-cylinder-compare-stressTT}
\end{figure}

We first consider the case without damage or fracture, focusing solely on the volume expansion of the reactive inner core and the resulting stress in the inert outer annulus.
As in the first example, the pore fluid pressure is kept at zero in both zones (\ie~$p_f = 0$).
Figures~\ref{fig:quarter-cylinder-compare-dispR} and \ref{fig:quarter-cylinder-compare-stressTT} compare the radial displacement ($u_{r}$) of the reaction zone and the angular hoop stress ($\stress_{\theta\theta}$) in the inert zone predicted by the two coupling methods, respectively.
As can be seen, the \revised{inelastic-strain-based} method predicts larger radial expansion of the inner core than the poromechanics-based method.
It is because the poromechanics-based method explicitly accounts for the compressibility of the pore mineral and fluid, which can accommodate part of the volumetric change through internal compression rather than transmitting it entirely as outward deformation.
As discussed in the analytical comparison under the zero-stress condition in Section~\ref{sec:analytical-comparison}, the poromechanics-based method yields the same volumetric expansion as the \revised{inelastic-strain-based} method only in the limit $\bar{K}^{\m} \gg K$ (\ie~incompressible mineral relative to the rock skeleton), otherwise the predicted expansion is smaller.
Regarding the stress induced in the inert zone, both methods capture the tensile hoop stress, indicating that they can both represent the tensile driving force for cracking associated with differential volume expansion in the confining medium.
However, the larger reactive-core expansion predicted by the \revised{inelastic-strain-based} method leads to a correspondingly larger stress magnitude.
Since pore minerals and fluids are compressible in reality, the \revised{inelastic-strain-based method} may overestimate the induced stress in the confined system and could therefore lead to an incorrect conclusion that tensile failure is imminent when it is not.
Therefore, selecting an appropriate chemo-mechanical coupling method is critical for correctly predicting the stress state in the host rock and informing reliable engineering decisions.

\begin{figure}[htbp]
    \centering
    \includegraphics[height=0.4\textheight]{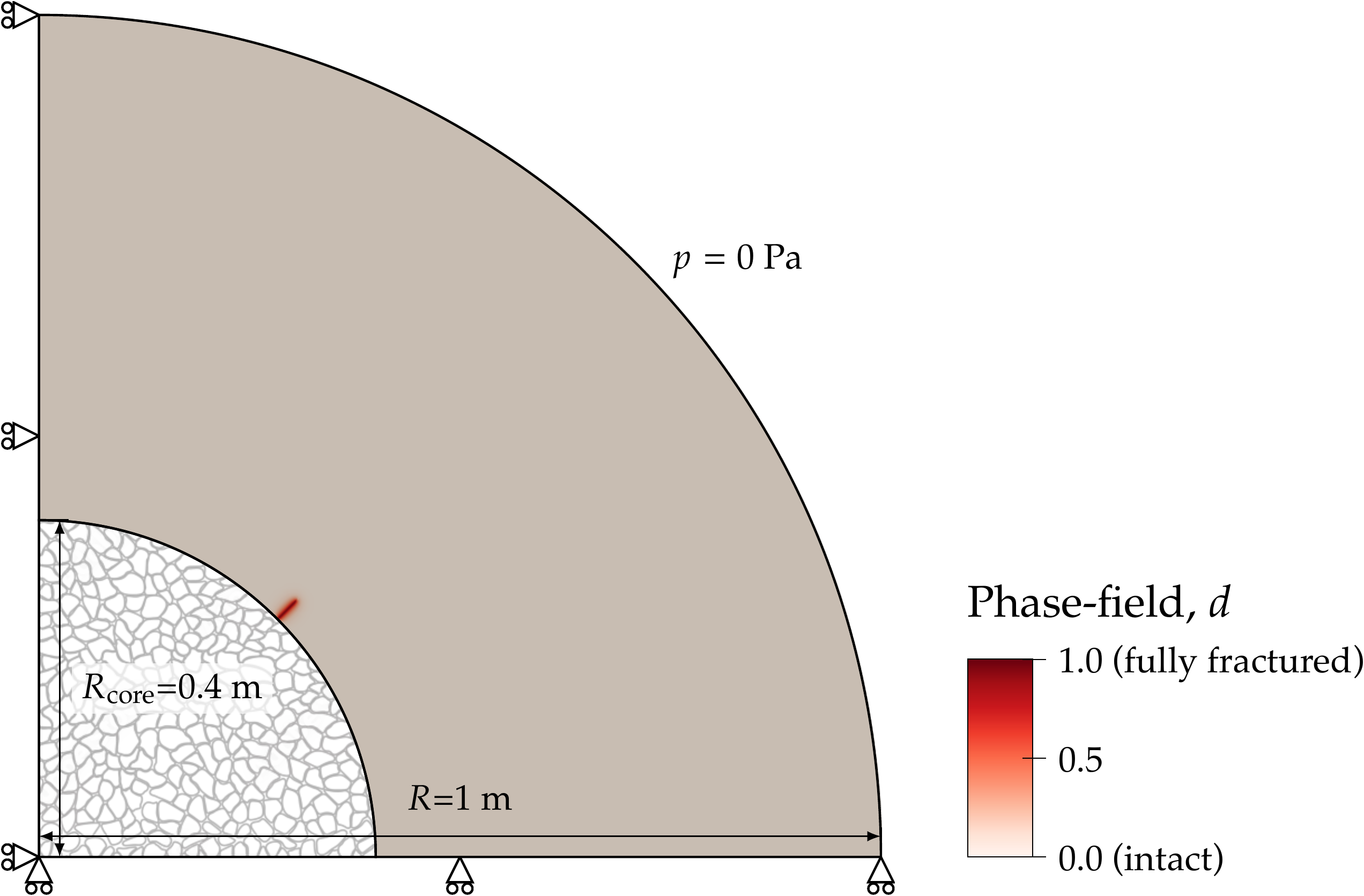}
    \caption{Serpentinization induced radial core expansion: problem setup with initial cracks.}
    \label{fig:quarter-cylinder-setup-cracks}
\end{figure}

We now extend the poromechanics-based approach to incorporate fracture, using the phase-field method to model crack propagation driven by mineralization-induced stress in the inert zone.
\revised{A separate inelastic-strain-based fracture simulation is omitted because both methods are expected to produce qualitatively similar fracture behavior, as both capture the mechanism described above in which reactive-core expansion induces tensile hoop stress in the inert annulus.}
As illustrated in Figure~\ref{fig:quarter-cylinder-setup-cracks}, an initial phase-field crack is assigned at $\theta = 45^\circ$ along the inner circumference of the inert zone to serve as a seed for fracture growth.
In essence, the phase-field method approximates a sharp fracture interface as a diffuse region characterized by a scalar damage variable $d \in [0, 1]$, where $d = 0$ indicates an intact material and $d = 1$ indicates complete fracture.
The detailed phase-field formulation adopted for this simulation is provided in~\ref{appendix:phasefield}.
The critical fracture energy and the phase-field length parameter are set to $\mathcal{G}_{c} = 0.1$ J/m$^2$ and $L_d = 0.005$ m, respectively.

To further investigate how mineralization-induced cracking promotes fluid flow and reactive transport in the host rock, we impose a fixed fluid pressure of $p_f = 1$~kPa in the reactive inner core, and a zero-pressure sink boundary condition at the outer circumference of the inert zone.
To account for the enhancement of permeability as fractures propagate, a simple damage-dependent model is adopted as a first-order approximation~\citep{pillai2018diffusive,mollaali2019numerical,fei2023phase},
\begin{align}
    k(d) = k_0 \exp(\beta_{k} d), \label{eq:damage-perm}
\end{align}
where $k_0$ is the reference permeability listed in Table~\ref{tab:quarter-cylinder-properties}, and $\beta_{k}$ is a damage-dependent scaling constant set to $\beta_{k} = 7$, following \citet{mollaali2019numerical,fei2023phase}.
Similarly, a damage-dependent diffusivity model of the same form is adopted,
\begin{align}
    \mathcal{D}(d) = \mathcal{D}_0 \exp(\beta_{D} d), \label{eq:damage-diff}
\end{align}
where $\mathcal{D}_0$ is the reference diffusivity coefficient in Table~\ref{tab:quarter-cylinder-properties}, and $\beta_{D}$ is a scaling constant set to $\beta_{D} = 13$.

\begin{figure}[htbp]
    \centering
    \includegraphics[width=0.9\textwidth]{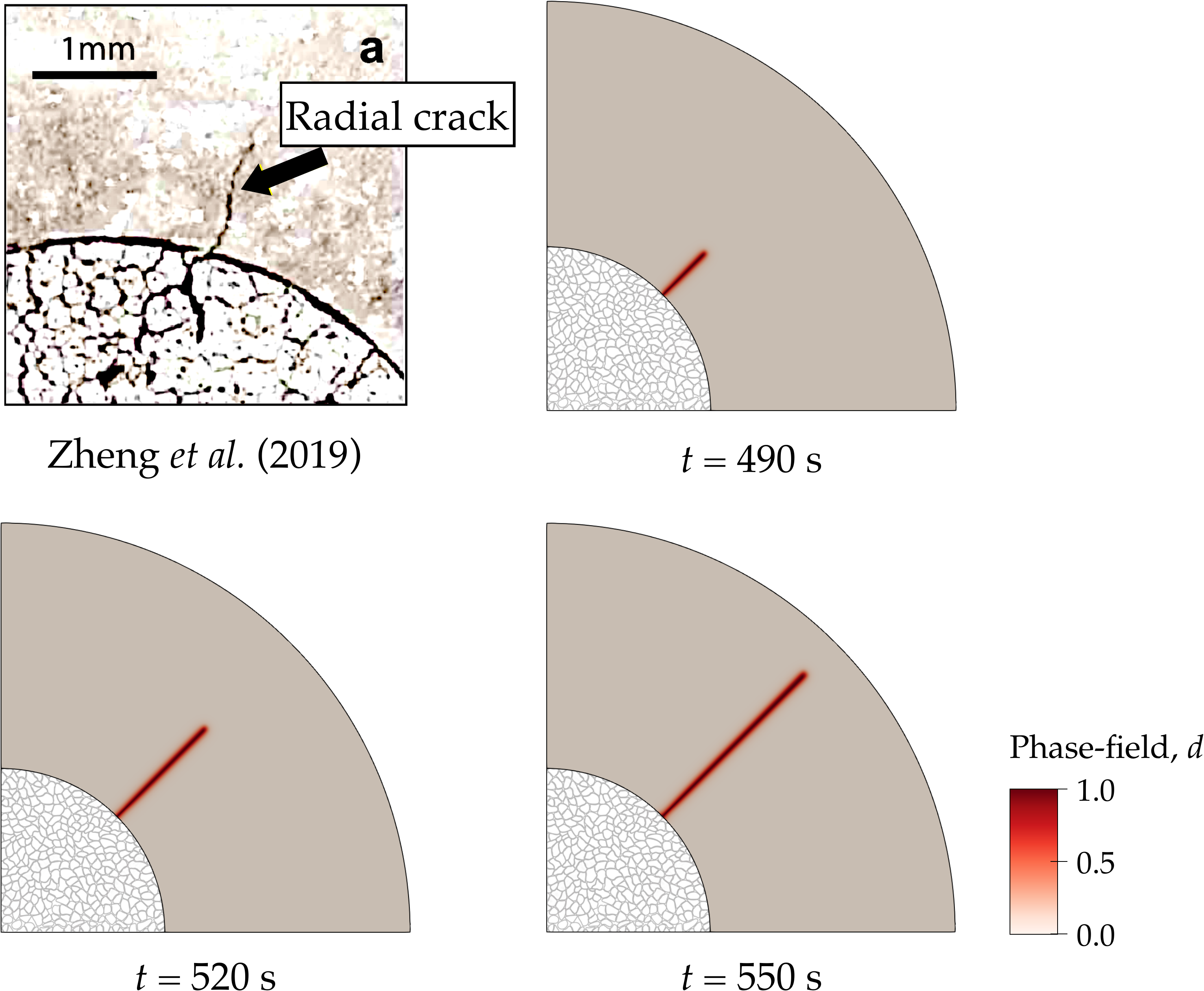}
    \caption{Serpentinization induced radial core expansion: phase-field fracture evolution in the inert zone, compared with the cracking pattern observed in the experiment~\citep{zheng2019mixed}.}
    \label{fig:quarter-cylinder-damage-cracks}
\end{figure}

\begin{figure}[htbp]
    \centering
    \includegraphics[height=0.2\textheight]{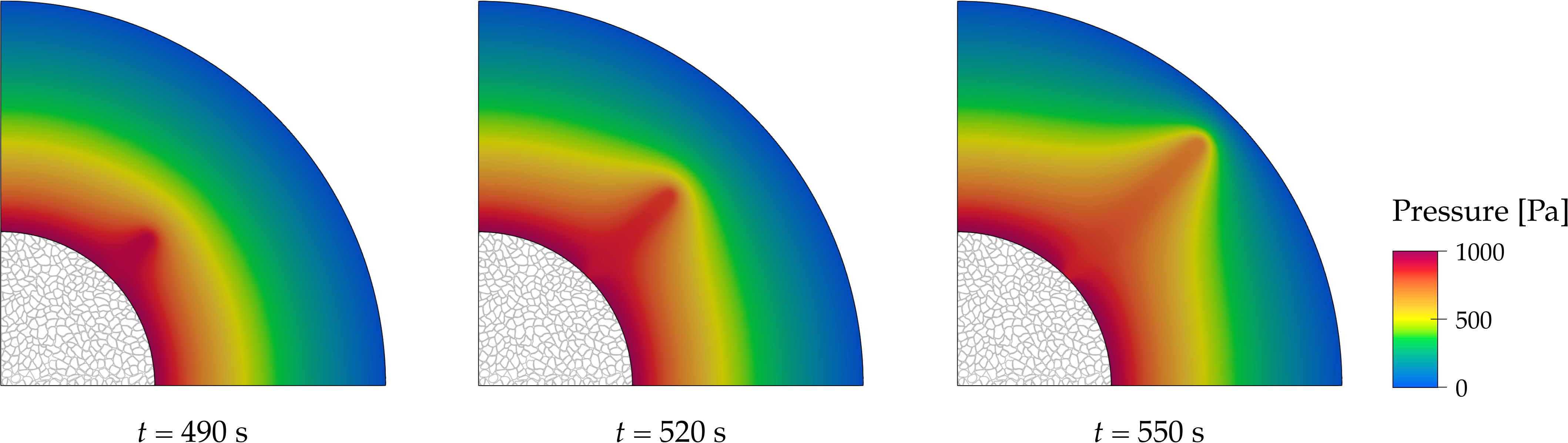}
    \caption{Serpentinization induced radial core expansion: evolution of fluid pressure in the inert zone.}
    \label{fig:quarter-cylinder-pres-cracks}
\end{figure}

\begin{figure}[htbp]
    \centering
    \includegraphics[width=\textwidth,height=0.2\textheight,keepaspectratio]{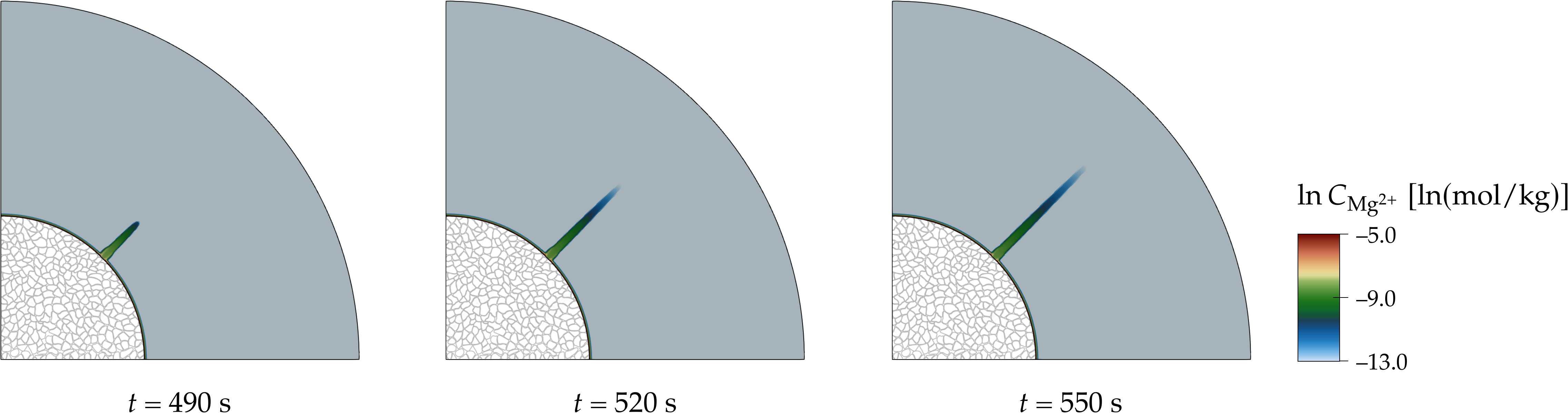}
    \caption{Serpentinization induced radial core expansion: evolution of $\ce{Mg^{2+}}$ concentration in the inert zone.}
    \label{fig:quarter-cylinder-lnMg-cracks}
\end{figure}

Figure~\ref{fig:quarter-cylinder-damage-cracks} presents the evolution of the phase-field fracture in the inert zone.
Driven by the tensile hoop stress induced by mineralization in the inner reactive core, the seed crack at $\theta = 45^\circ$ propagates radially outward with time.
Also notably, this radial propagation pattern is qualitatively consistent with the cracking behavior observed in laboratory experiments on reaction-driven fracturing~\citep{zheng2019mixed} as shown in the top left panel of the figure.
Following the fracture development, Figures~\ref{fig:quarter-cylinder-pres-cracks} and \ref{fig:quarter-cylinder-lnMg-cracks} present the corresponding evolution of the fluid pressure and $\ce{Mg^{2+}}$ concentration in the inert zone.
Both the pressure and $\ce{Mg^{2+}}$ concentration fronts migrate radially outward along the fracture path, indicating that the mineralization-induced fracturing locally enhances fluid flow and reactive transport.
This amplified flow and species transport could, in turn, drive additional reactions in regions further away from the reactive core (although such far-field reactions are not modeled here), suggesting a potentially self-accelerating process with important implications for controlling fluid flow in subsurface energy applications.

\subsection{Carbonation-induced fracture development from pore-scale mineral growth}

The third example aims to demonstrate the unique capability of the poromechanics-based method to bridge pore-scale mineral growth and the resulting fracture development in rocks at the continuum scale.
For this purpose, we consider a carbonation reaction to represent the scenario of mineralization at the pore and potentially induced cracking due to supercritical \ce{CO2} injection, 
\begin{align}
    \ce{CaCO3 (calcite) + H+ &<=> Ca^{2+} + HCO3-}. \label{eq:calcite}
\end{align}
The kinetic rate constant and equilibrium constant for this reaction are listed in Table~\ref{tab:carbonation-properties}.

\begin{table}[htbp]
\caption{Kinetic and thermodynamic parameters for the carbonation reaction.}
\label{tab:carbonation-properties}
\centering
\begin{tabular}{p{4.0cm}ccc}
\toprule
\textbf{Property} & \textbf{Symbol} & \textbf{Calcite} & \textbf{Unit} \\
\midrule
Rate constant      & $k_m$               & $1.47\times 10^{-6}$ & mol\,m$^{-2}$\,s$^{-1}$ \\
Equilibrium constant & $K^\mathrm{eq}_{m}$ & $49$ & {--} \\
\bottomrule
\end{tabular}
\end{table}

\begin{figure}[htbp]
    \centering
    \includegraphics[width=\textwidth]{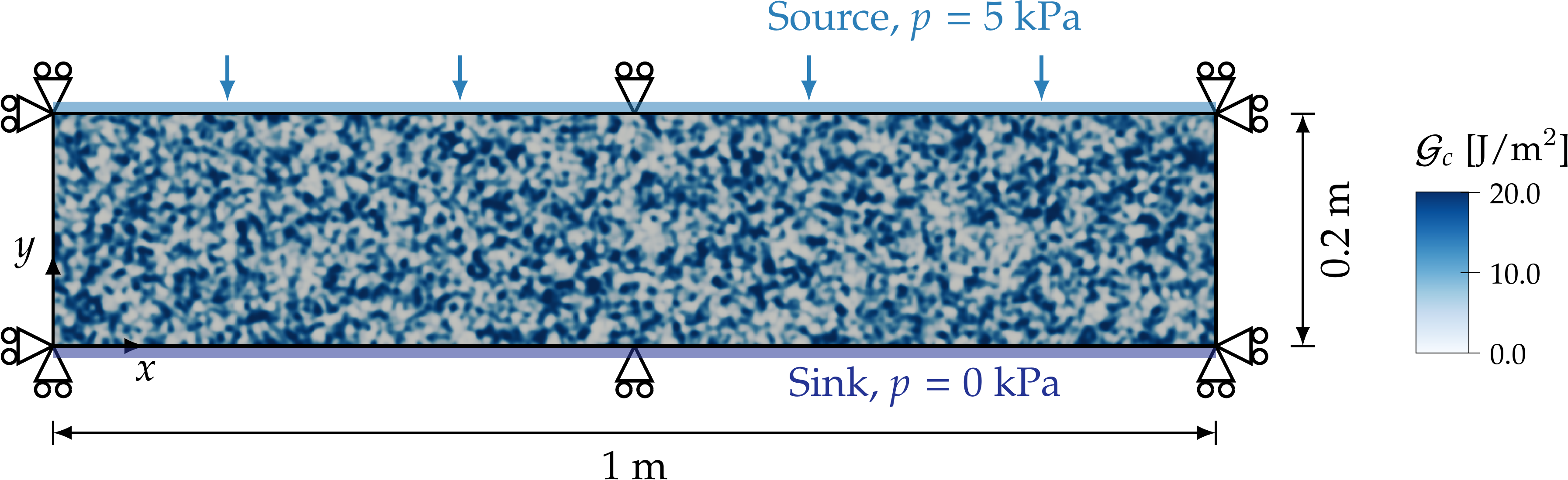}
    \caption{Carbonation-induced fracture development from pore-scale mineral growth: problem setup.}
    \label{fig:pore-crack-setup}
\end{figure}

Figure~\ref{fig:pore-crack-setup} illustrates the problem setup, where the domain is a rectangular strip of 1~m in length and 0.2~m in width.
For the mechanical boundary conditions, rollers are applied along all four boundaries, such that the domain is fully confined to induce stress arising from mineralization.
For the flow and geochemistry, a fluid source with a fixed pressure of 5~kPa and high \ce{HCO3-} concentration is imposed along the top boundary, representing the influx of a carbonate-rich solution associated with \ce{CO2} injection, while the bottom boundary serves as a zero-pressure sink.
The initial species concentrations in the domain and those fixed at the source are summarized in Table~\ref{tab:pore-crack-concentrations}.
The initial concentrations are chosen such that the reaction rate is zero at the beginning according to Eq.~\eqref{eq:reaction-rate}, \ie~the pore fluid is initially in equilibrium with calcite.
As the injected fluid infiltrates the domain, the elevated \ce{HCO3-} concentration drives Eq.~\eqref{eq:calcite} in the reverse direction, precipitating calcite in the pore space, and the resulting pore mineral pressure loads the surrounding rock skeleton.

\begin{table}[htbp]
\caption{Carbonation-induced fracture development from pore-scale mineral growth: initial species concentrations in the domain and the fixed concentrations at the source.}
\label{tab:pore-crack-concentrations}
\centering
\begin{tabular}{p{5.0cm}cccc}
\toprule
\textbf{Species} & \textbf{Symbol} & \textbf{Domain (initial)} & \textbf{Source (fixed)} & \textbf{Unit} \\
\midrule
Hydrogen ion, \ce{H+}          & $C_{\mathrm{H^+}}$      & $10^{-8}$          & $10^{-8}$          & mol/kg \\
Calcium ion, \ce{Ca^{2+}}      & $C_{\mathrm{Ca^{2+}}}$  & $7\times10^{-4}$   & $7\times10^{-4}$   & mol/kg \\
Bicarbonate ion, \ce{HCO3-}    & $C_{\mathrm{HCO_3^-}}$  & $7\times10^{-4}$   & $5\times10^{-2}$   & mol/kg \\
\bottomrule
\end{tabular}
\end{table}

The material properties adopted in this example are summarized in Table~\ref{tab:pore-crack-properties}, where the damage-dependent permeability and diffusivity models in Eqs.~\eqref{eq:damage-perm} and \eqref{eq:damage-diff} are again adopted.
To reflect the structural heterogeneity of the rock at the continuum scale, the critical fracture energy $\mathcal{G}_{c}$ is prescribed as a spatially heterogeneous field ranging from 0.01 to 20~J/m$^2$, as depicted in Figure~\ref{fig:pore-crack-setup}.
The low-$\mathcal{G}_{c}$ regions represent weak zones where fractures preferentially develop, thereby allowing the simulation to capture crack nucleation and propagation driven by pore-scale mineral growth.

\begin{table}[htbp]
\caption{Carbonation-induced fracture development from pore-scale mineral growth: material properties. Properties marked with $(\dagger)$ are specific to the poromechanics-based coupling model.}
\label{tab:pore-crack-properties}
\centering
\begin{tabular}{p{2.3cm}p{5.0cm}ccc}
\toprule
\multicolumn{1}{l}{\textbf{Category}} &
\multicolumn{1}{l}{\textbf{Property}} &
\textbf{Symbol} & \textbf{Value} &
\multicolumn{1}{c}{\textbf{Unit}} \\
\midrule
\multirow{9}{*}{\parbox{2.3cm}{Rock matrix}}
 & Drained bulk modulus        & $K$   & 6.67 & GPa \\
 & Shear modulus               & $G$   & 100 & GPa \\
 & Initial porosity            & $\phi_0$            & 0.1 & --  \\
 & Solid grain bulk modulus$^{\dagger}$ & $K^{\mathrm{s}}$      & $10^{20}$ & GPa \\
 & Reference permeability      & $k_0$               & $3\times 10^{-15}$ & m$^2$ \\
 & Permeability scaling constant & $\beta_{k}$       & 7.0 & -- \\
 & Reference diffusivity       & $\mathcal{D}_{0}$   & $10^{-6}$ & m$^2$/s \\
 & Diffusivity scaling constant & $\beta_{D}$        & 10.0 & -- \\
 & Phase-field length parameter & $L_d$              & 0.002 & m \\
\midrule
\multirow{2}{*}{\parbox{2.3cm}{Fluid}}
 & Dynamic viscosity    & $\mu_f$    & $10^{-3}$ & Pa$\cdot$s \\
 & Compressibility      & $c_f$      & $5 \times 10^{-10}$ & Pa$^{-1}$ \\
\midrule
\multirow{5}{*}{\parbox{2.3cm}{Mineral \\ (calcite)}}
 & Mineral bulk modulus$^{\dagger}$ & $\bar{K}^{\m}$          & 6.67 & GPa \\
 & Initial volume fraction          & $\theta_{\mathrm{Cal},0}$ & 0.02 & -- \\
 & Molar weight                     & $M_{\mathrm{Cal}}$      & 100.09 & g/mol \\
 & Mineral density                  & $\rho_{\mathrm{Cal}}$   & 2710.0 & kg/m$^3$ \\
 & Initial surface area             & $A_{\mathrm{Cal},0}$ & 0.1 & m$^2$/m$^3$ \\
\bottomrule
\end{tabular}
\end{table}

\begin{figure}[htbp]
    \centering
    \includegraphics[height=0.28\textheight]{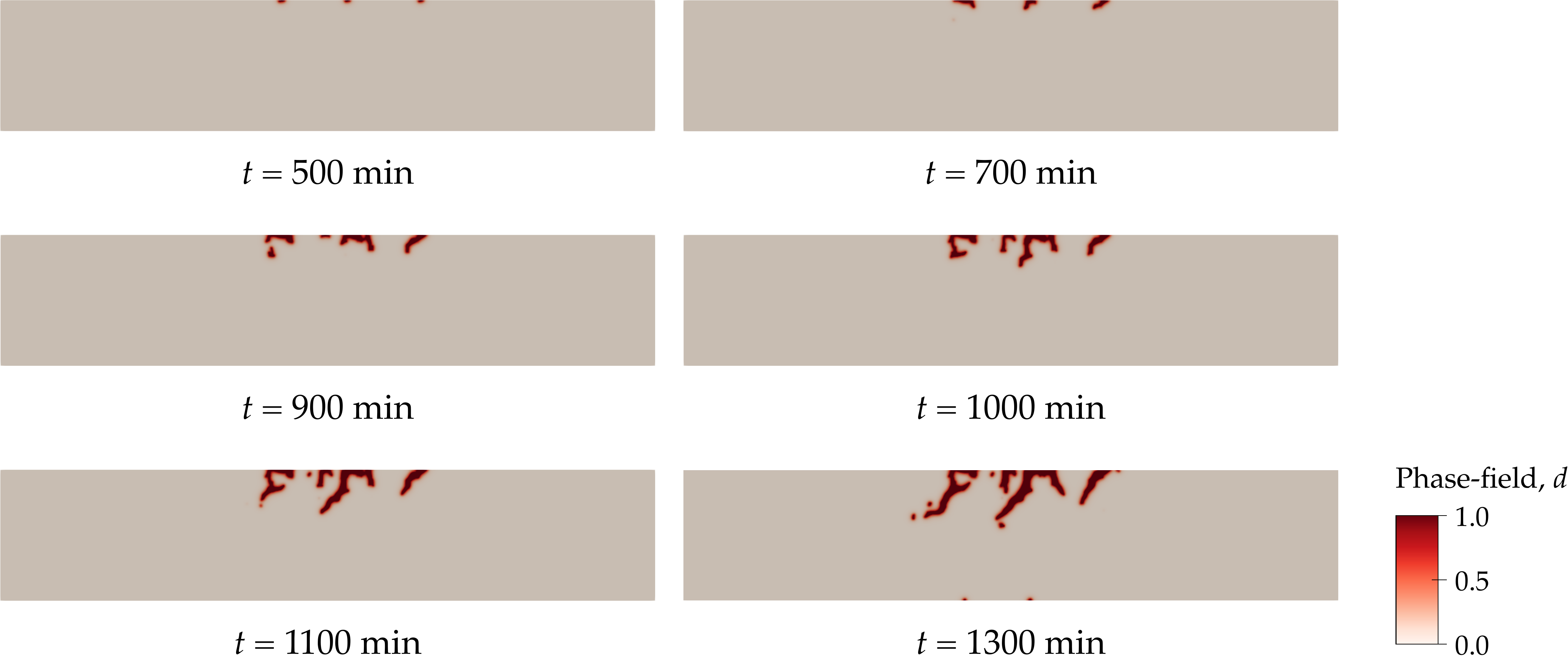}
    \caption{Carbonation-induced fracture development from pore-scale mineral growth: phase-field fracture evolution modeled by the poromechanics-based method.}
    \label{fig:pore-crack-damage}
\end{figure}

Figure~\ref{fig:pore-crack-damage} presents the evolution of the phase-field fracture predicted by the poromechanics-based method.
As the carbonate-rich fluid infiltrates downward from the source, calcite precipitates in the pore space of the upper portion of the strip, where the mineral volume growth generates additional pore pressure that exerts a splitting force on the hosting rock skeleton.
As can be seen, damage begins to nucleate at several weak sites (low-$\mathcal{G}_{c}$) beneath the top boundary at around $t = 500$~min, even though no initial crack is prescribed anywhere in the domain.
As precipitation continues, the cracks propagate and penetrate deeper into the strip.
This crack nucleation-propagation process demonstrates that the poromechanics-based framework can naturally bridge pore-scale mineral growth and continuum-scale fracture development, thereby enabling direct investigation of the mineralization-pressure mechanism of reaction-induced fracturing.

\begin{figure}[htbp]
    \centering
    \subfloat[]{\includegraphics[width=\textwidth]{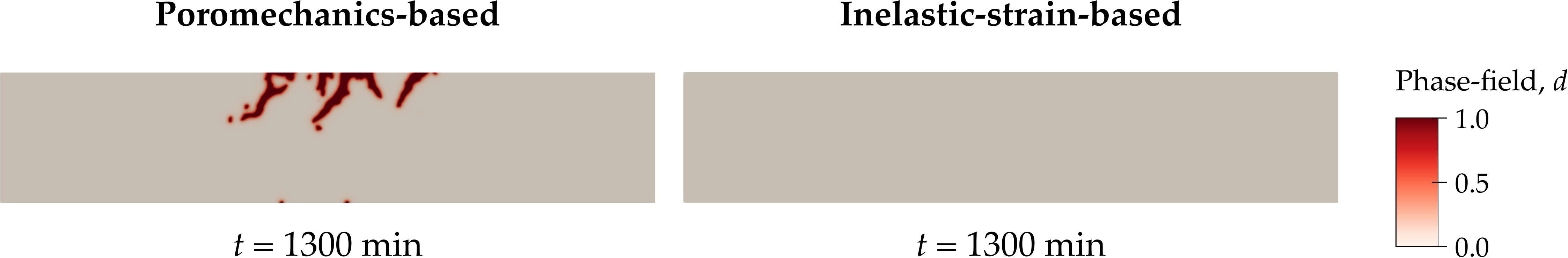}\label{fig:pore-crack-damage-comparison}}
    \\
    \subfloat[\revised{Poromechanics-based}]{\includegraphics[width=0.48\textwidth]{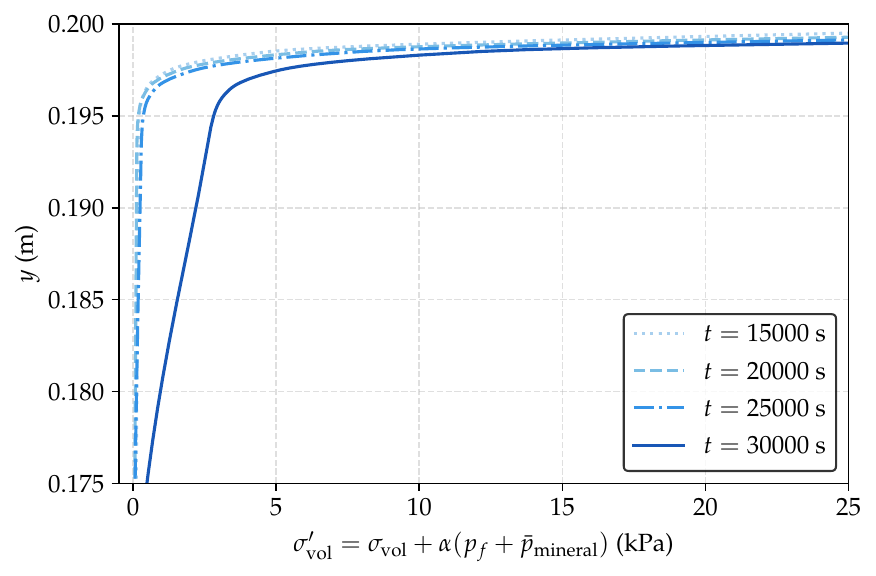}\label{fig:pore-crack-poromech-meanStress}}
    \hfill
    \subfloat[\revised{Inelastic-strain-based}]{\includegraphics[width=0.48\textwidth]{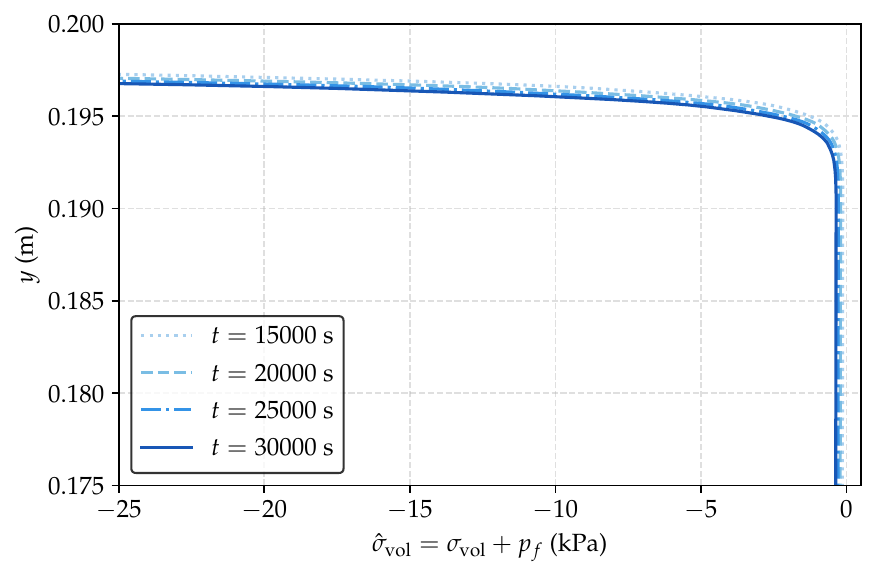}\label{fig:pore-crack-eigenstrain-meanStress}}
    \caption{Carbonation-induced fracture development from pore-scale mineral growth: (a) comparison of the phase-field damage predicted by the poromechanics-based and \revised{inelastic-strain-based} methods; (b) volumetric effective stress profiles predicted by the poromechanics-based method; (c) volumetric effective stress profiles predicted by the \revised{inelastic-strain-based} method.}
    \label{fig:pore-crack-eigen-damage}
\end{figure}

By contrast, the \revised{inelastic-strain-based} method predicts no damage development at the end of the simulation, as shown in Figure~\ref{fig:pore-crack-damage-comparison}.
This contrast is rooted in how the two methods translate mineral growth into the mechanical stress field, as also briefly mentioned in Section~\ref{sec:1d-serpentinization}.
\revised{More specifically, the poromechanics-based method formulates the pore mineral pressure $\bar{p}_{\m}$ separately from the host-rock effective stress, allowing mineral pressure to be explicitly captured as a distinct mechanical loading on the pore walls that drives the evolution of the effective stress.}
As shown in Figure~\ref{fig:pore-crack-poromech-meanStress}, the resulting volumetric effective stress of the host rock, $\stress'_{\vol}$, is tensile and intensifies as precipitation proceeds until the associated strain energy is sufficient to cause damage.
In the \revised{inelastic-strain-based} method, however, calcite precipitation enters the formulation as an inelastic expansion of the entire bulk solid, which is fully accommodated by elastic compression under the confined condition.
This is evidenced by Figure~\ref{fig:pore-crack-eigenstrain-meanStress}, where the lumped volumetric effective stress of the host rock and solid minerals, $\hat{\stress}_{\vol}$, remains compressive near the source, providing no tensile driving force for the phase field.
Overall, the poromechanics-based method intrinsically accounts for the microstructural behavior at the macroscopic scale by distinguishing among the host-rock effective stress, the pore mineral pressure, and the pore fluid pressure.
As such, the mechanical response and the resulting damage or fracture can be simulated at the continuum scale without explicitly representing the microstructural features, which is particularly beneficial for large, reservoir-scale simulations.

\begin{figure}[htbp]
    \centering
    \includegraphics[height=0.28\textheight]{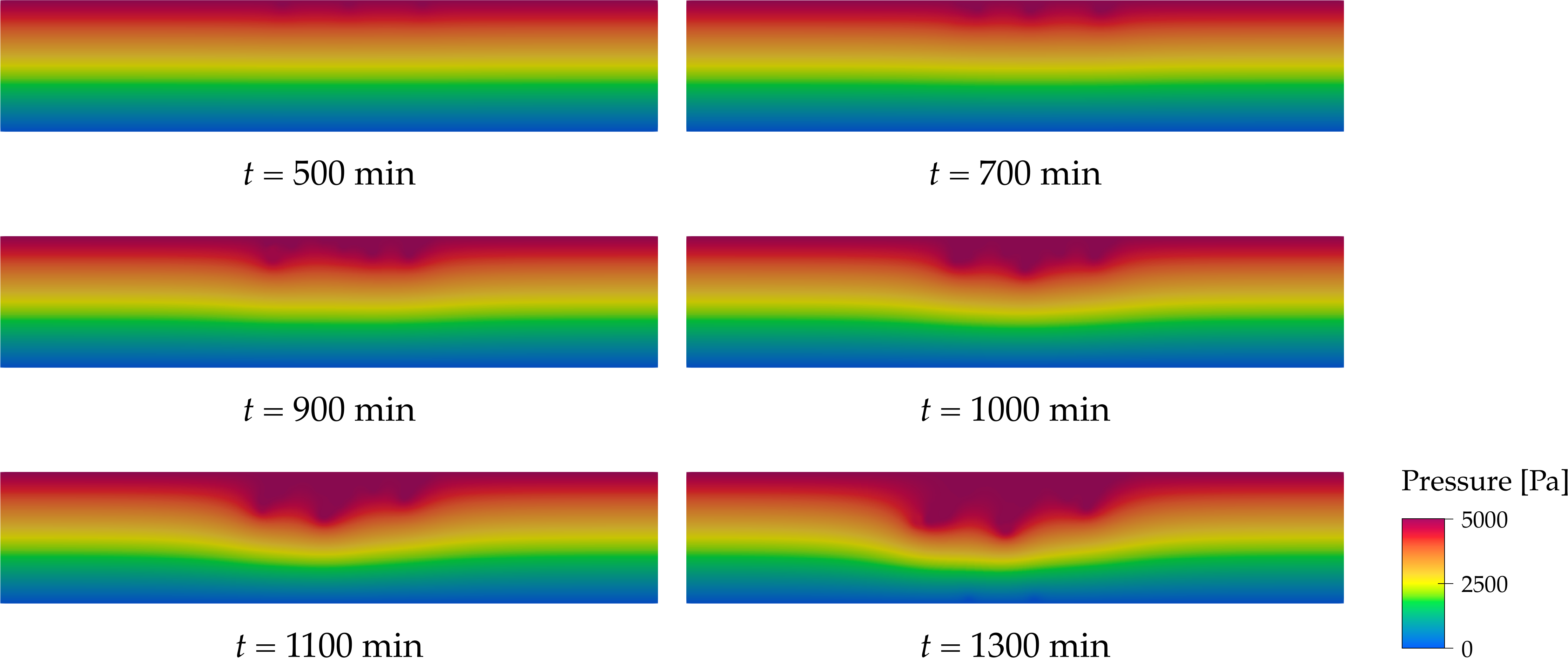}
    \caption{Carbonation-induced fracture development from pore-scale mineral growth: evolution of the fluid pressure modeled by the poromechanics-based method.}
    \label{fig:pore-crack-pres}
\end{figure}

\begin{figure}[htbp]
    \centering
    \includegraphics[height=0.28\textheight]{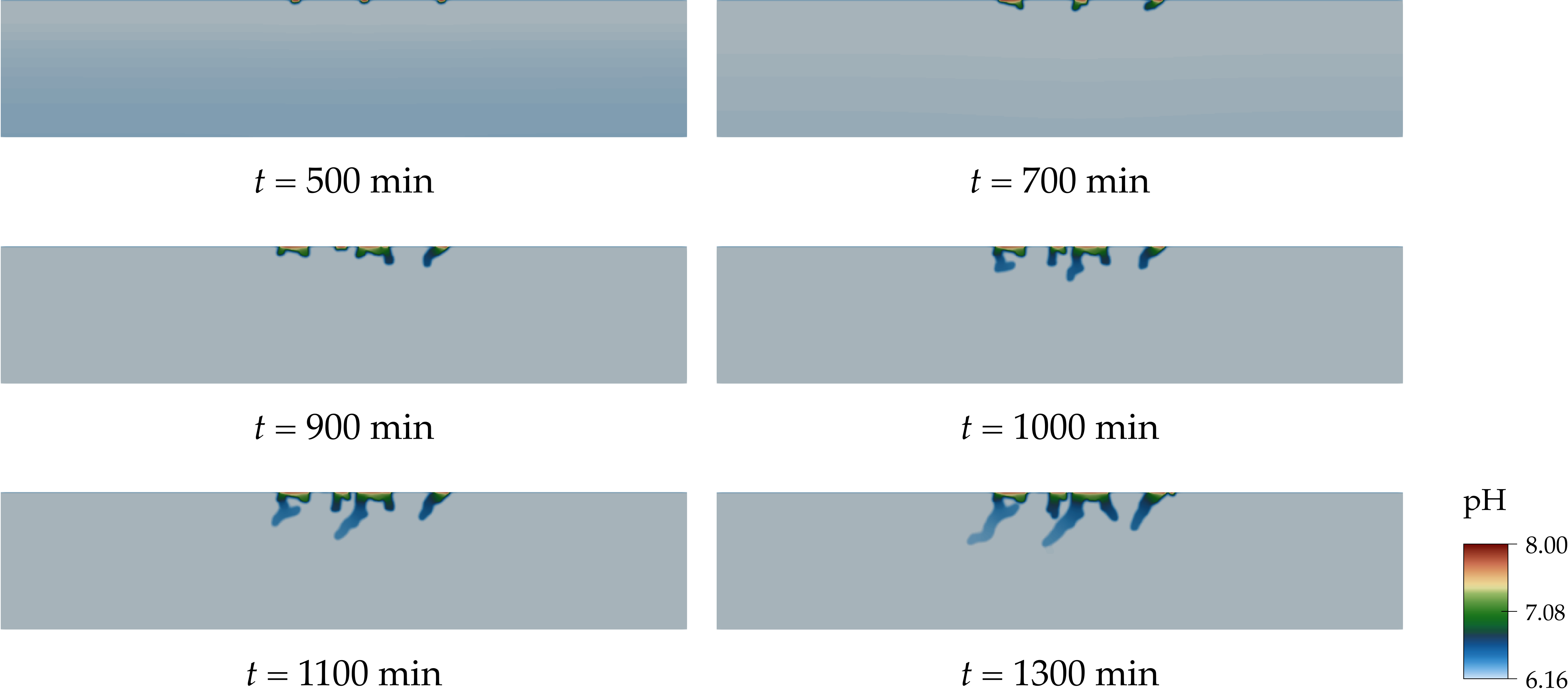}
    \caption{Carbonation-induced fracture development from pore-scale mineral growth: evolution of pH modeled by the poromechanics-based method.}
    \label{fig:pore-crack-concH}
\end{figure}

\begin{figure}[htbp]
    \centering
    \includegraphics[height=0.28\textheight]{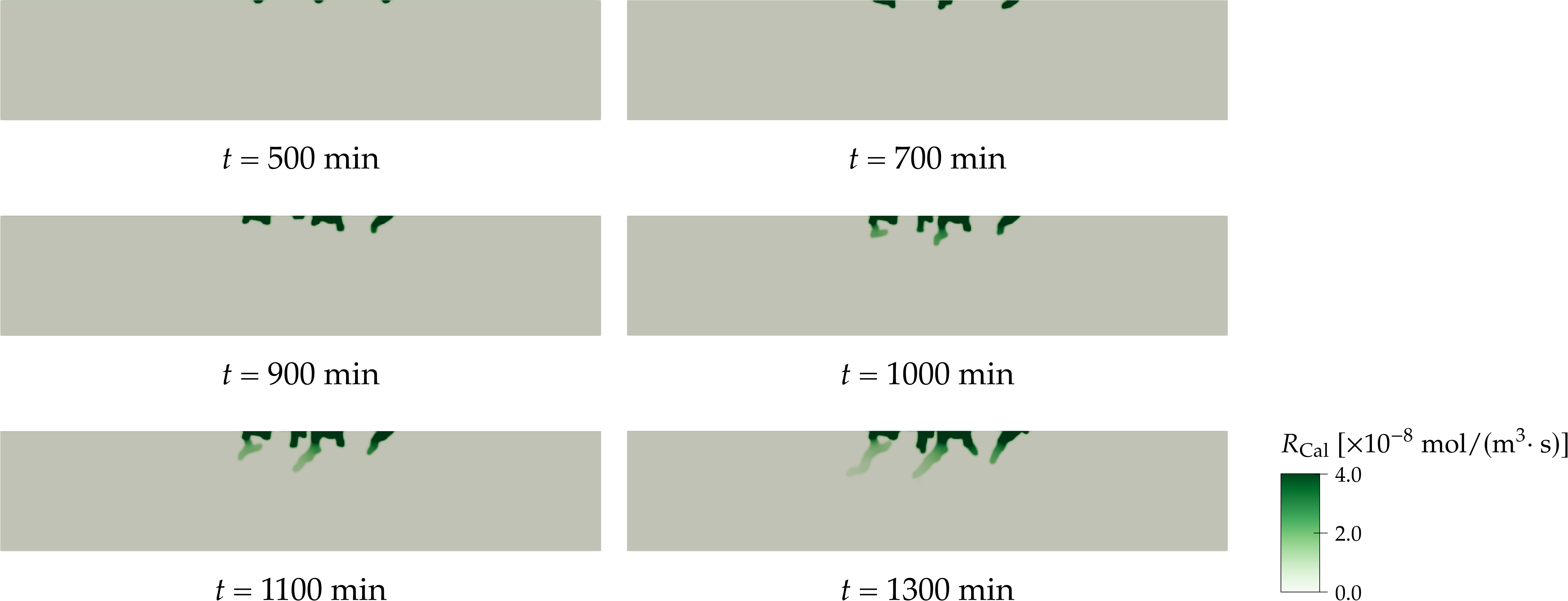}
    \caption{Carbonation-induced fracture development from pore-scale mineral growth: evolution of the calcite reaction rate modeled by the poromechanics-based method.}
    \label{fig:pore-crack-reactionRate}
\end{figure}

We next examine the flow and chemistry results of the poromechanics-based method to elucidate the coupled reactive transport processes during the fracture development.
Figure~\ref{fig:pore-crack-pres} shows the evolution of the fluid pressure, where the enhanced permeability of the damaged zone allows the high-pressure front to advance faster along the fracture network, accelerating the advection of \ce{HCO3-}-rich fluid.
Figures~\ref{fig:pore-crack-concH} and \ref{fig:pore-crack-reactionRate} present the corresponding evolutions of pH (\ie~$- \log_{10}(C_{\ce{H+}})$) and the kinetic reaction rate of calcite, respectively.
Since calcite precipitation releases hydrogen ions (see Eq.~\eqref{eq:calcite}), the pH decreases from its initial value of 8 throughout the domain as the reaction proceeds.
Meanwhile, the fractured region maintains a slightly higher pH, because the enhanced diffusion and advection continuously deliver high-pH fluid from the source.
Additionally, as fractures develop, the reaction front extends deeper along the cracks, where the enhanced transport continuously replenishes \ce{HCO3-}.
Similar to the second example, these results reveal a positive feedback loop among mineral precipitation, fracturing, and reactive transport.

\section{Closure}
\label{sec:closure}


This work has presented a poromechanics-based approach to chemo-mechanics coupling and integrated it into coupled reactive transport and geomechanics simulations of geochemical mineralization in porous rocks.
Drawing from the classical poromechanics framework, the formulation explicitly distinguishes among the effective stress in the host rock, the pore fluid pressure, and the pore mineral pressure associated with precipitation and dissolution.
This treatment provides a generalized, mathematically tractable, and physically grounded model for chemo-mechanical coupling, with a porosity update that accounts for pore structure deformation and the compressibility of pore fluids and minerals.
The resulting chemo-hydro-mechanical model has been implemented in GEOS using a sequential scheme between finite-volume reactive transport and finite-element geomechanics.

\revised{Through numerical comparisons}, we highlighted the unique capabilities of the poromechanics-based approach for capturing the mechanical consequences of pore-scale mineral growth.
In particular, by accounting for mineral compressibility and its interactions with the host-rock pore structure and fluid, the poromechanics-based method translates pore-scale mineral growth into pore mineral pressure as a separate mechanical driver.
This capability provides a more flexible means of modeling reaction-induced volume changes, effective stress evolution, and fracture development at the continuum scale without explicitly resolving the pore geometry and other microstructural features.
With damage-dependent permeability and diffusivity, the examples further demonstrated that mineralization-induced fracturing and damage can enhance fluid flow and species transport, suggesting a feedback between reactions, mechanics, and transport.

Future work will focus on additional experimental validation and improved coupling with fracture and evolving geochemical and transport properties.
With advanced modern computing capabilities, the implementation in GEOS also provides a foundation for three-dimensional, reservoir-scale simulations of reactive subsurface systems relevant to carbon storage, geothermal energy, and related applications~\citep{cusini2023simulation,camargo2024managing,huang2026simulation,kroll2026coupled,zibitsker2026basin}.

\section*{Acknowledgements}
This work was performed under the auspices of the U.S. Department of Energy by Lawrence Livermore National Laboratory under Contract DE-AC52-07NA27344.
Fei, Zibitsker, Settgast, Finstad, and Bucci acknowledge the support by the project ``Electro-chemo-mechanics of reactive interfaces to achieve energy and
carbon storage at scale'' funded by the Laboratory Directed Research and Development (LDRD) Program at Lawrence Livermore National Laboratory under award \#25-ERD-006.
Yang and Buscarnera acknowledge the support by the Center on Geo-process in Mineral Carbon Storage, an Energy Frontier Research Center (EFRC) funded by the U.S. Department of Energy, Office of Science, Basic Energy Sciences at the University of Minnesota under award \#DE-SC0023429.
The authors also wish to thank the GEOS development team for their contributions.

\section*{CRediT authorship contribution statement}
\textbf{Fan Fei}: Conceptualization, Methodology, Software, Validation, Formal analysis, Investigation, Data curation, Visualization, Writing - original draft, Writing - review \& editing.
\textbf{Yifan Yang}: Conceptualization, Methodology, Validation, Formal analysis, Investigation, Writing - original draft, Writing - review \& editing.
\textbf{Aleksander Zibitsker}: Software, Validation, Investigation, Writing - review \& editing.
\textbf{Giuseppe Buscarnera}: Conceptualization, Methodology, Funding acquisition, Supervision, Writing - review \& editing.
\textbf{Randolph Settgast}: Software, Resources, Writing - review \& editing.
\textbf{Kari Finstad}: Funding acquisition, Project administration, Supervision, Resources, Writing - review \& editing.
\textbf{Giovanna Bucci}: Funding acquisition, Project administration, Supervision, Resources, Writing - review \& editing.

\section*{Declaration of generative AI use}
During the preparation of the manuscript, the authors used OpenAI's ChatGPT/Codex and Anthropic's Claude to assist with language editing and drafting suggestions.
After using this tool, the authors reviewed and edited the content as needed and take full responsibility for the content of the publication.

\section*{Declaration of competing interest}
The authors declare that they have no known competing financial interests or personal relationships that could have appeared to influence the work reported in this paper.

\section*{Data availability statement}
The code used in this study is available in the GEOS GitHub repository: \url{https://github.com/GEOS-DEV}.
Simulation input files and data are available from the corresponding authors upon reasonable request.

\appendix

\section{Phase-field modeling of fracture}
\label{appendix:phasefield}

This appendix briefly introduces the phase-field method and presents its formulations for modeling mineralization-induced fracturing in the numerical examples.

The phase-field method approximates a sharp fracture interface as a diffuse region characterized by a scalar variable $d \in [0, 1]$, where $d = 0$ represents the intact material and $d = 1$ represents complete fracture.
To account for the degradation of material stiffness upon fracturing, the effective stress constitutive relation in Eq.~\eqref{eq:biot-total-stress} is modified as
\begin{align}
    \dot{\stress}'_{ij} = g(d)\, \mathbb{C}_{ijkl} \dot{\strain}_{kl},
\end{align}
where $g(d)$ is the degradation function, which should satisfy
\begin{align}
    g(0) = 1, \, g(1) = 0, \, g'(1) = 0, \,\, \text{and} \,\, g'(d) < 0 \,\, \text{for} \,\, d\in [0, 1).
\end{align} 
While various stress decomposition schemes have been developed~\citep{amor2009regularized,miehe2010thermodynamically,steinke2019phase,fei2020phaseb,fei2021double}, we adopt full degradation of the stiffness tensor for simplicity.

Fracture propagation is described by the evolution of the phase-field variable, following the governing equation below derived from the fracture mechanics principle,
\begin{align}
    g'(d) \mathcal{H}^{+} + \dfrac{\mathcal{G}_{c}}{c_0 L_d } \left[ \omega'(d) - 2L^2_d \dfrac{\pd^2 d}{\pd x_i \pd x_i} \right] = 0,  \quad \text{with} \quad c_{0} = 4\int_{0}^{1} \sqrt{\omega(s)} \: \dd s. \label{eq:phase-field_generic}
\end{align}
Here, $\mathcal{H}^{+} = \max_{\theta \in[0, t]} \psi^{\el}\!\left(\strain_{ij}(\theta)\right)$ is a history variable of crack driving force that stores the maximum strain energy density, $\mathcal{G}_c$ is the critical energy release rate, $L_d$ is the phase-field length scale that controls the width of the diffuse crack, and $\omega(d)$ is the local dissipation function.
We adopt the standard quadratic degradation function, \ie~$g(d) = (1-d)^2$, together with the linear dissipation function $\omega(d) = d$ that defines the \texttt{AT1} model~\citep{pham2011gradient,geelen2019phase,kumar2020revisiting,fei2023phaseb}, which gives $c_0 = 8/3$.
In the \texttt{AT1} model, damage growth is activated only when the strain energy density exceeds the critical threshold $\psi_c = 3\mathcal{G}_c/(16L_d)$.
To ensure a non-negative phase-field value derived from Eq.~\eqref{eq:phase-field_generic}, we enforce $\mathcal{H}^{+} \geq \psi_c$, thus
\begin{align}
    \mathcal{H}^{+} = \max_{\theta \in[0, t]} \left[ \psi^{\el}\!\left(\strain_{ij}(\theta)\right), \, \psi_c \right] \label{eq:history-variable}.
\end{align}

To further account for the tension-compression asymmetry in fracture propagation, we evaluate $\mathcal{H}^{+}$ as the maximum positive part of the strain energy density following the spectral decomposition of the strain energy density~\citep{miehe2010thermodynamically}.
Combining Eq.~\eqref{eq:history-variable}, the final form of the crack driving force is given by
\begin{align}
    \mathcal{H}^{+} = \max_{\theta \in[0, t]} \left [ \psi^{+}\!\left(\strain_{ij}(\theta)\right), \, \psi_c \right ],
\end{align}
where $\psi^{\pm}$ is the tensile/compressive part of the strain energy density given by
\begin{align}
    \psi^{\pm}(\strain_{ij}) = \frac{\lambda}{2} \langle \strain_{1} + \strain_{2} + \strain_{3} \rangle_{\pm}^2 + G \left( \langle \strain_{1} \rangle_{\pm}^2 + \langle \strain_{2} \rangle_{\pm}^2 + \langle \strain_{3} \rangle_{\pm}^2 \right),
\end{align}
where $\lambda$ is the first Lamé parameter, $\strain_{1}$, $\strain_{2}$, and $\strain_{3}$ are the principal strains, and $\langle \cdot \rangle_{\pm}$ denotes the positive/negative part of a scalar, \ie
\begin{align}
    \langle x \rangle_{+} = \max(x,\, 0), \quad \langle x \rangle_{-} = \min(x,\, 0).
\end{align}
It is noted that when the \revised{inelastic-strain-based} method is used, the strain $\strain_{ij}$ in the above formulation refers to the mechanical elastic strain only, \ie~$\strain^{\el}_{ij} = \strain_{ij} - \strain^{\ast}_{ij}$, excluding the \revised{inelastic strain} due to mineral growth.


\bibliography{references}

\end{document}